\documentclass[11pt,fleqn]{article}
\usepackage{amssymb}
\usepackage{graphicx}
\usepackage{amssymb,latexsym,amsmath,amsfonts,graphicx, ebezier}

    \advance\textheight by \topskip

    \numberwithin{equation}{section}

    \def\Re{{\rm Re \,}}
    \def\Im{{\rm Im \,}}

    \def\bigO{{\cal O}}
    
    \def\Res{{\rm Res}}

\newcommand{\e}{\epsilon}
\newcommand{\lb}{\lambda}

    \newtheorem{theorem}{Theorem}[section]

    \newtheorem{Definition}[theorem]{Definition}
    
    \newtheorem{Remark}[theorem]{Remark}
    \newenvironment{remark}{\begin{Remark}\rm}{\end{Remark}}
    \newtheorem{Example}[theorem]{Example}
    
    \newtheorem{Assumptions}[theorem]{Assumptions}

    {\rm \trivlist \item[\hskip \labelsep{\bf Proof. }]}%
    {\hspace*{\fill}$\Box$\endtrivlist}
    {\rm \trivlist \item[\hskip \labelsep{\bf Proof}]}%
    {\hspace*{\fill}$\Box$\endtrivlist}

\begin{document}
\title{The KdV hierarchy: universality and a Painlev\'e transcendent}
\author{T. Claeys and T. Grava}

\maketitle

\begin{abstract}
We study the Cauchy problem for the Korteweg-de Vries (KdV)
hierarchy in the small dispersion limit where $\e\to 0$. For
negative analytic initial data with a single negative hump, we prove
that for small times, the solution is approximated by the solution
to the hyperbolic transport equation which corresponds to $\e=0$.
Near the time of gradient catastrophe for the transport equation, we
show that the solution to the KdV hierarchy is approximated by a
particular Painlev\'e transcendent. This supports Dubrovins
universality conjecture concerning the critical behavior of
Hamiltonian perturbations of hyperbolic equations. We use the
Riemann-Hilbert approach to prove our results.
\end{abstract}

\section{Introduction}
In this manuscript, we will prove a particular case of a conjecture
in \cite{dubcr} about the formation of dispersive shocks \cite{GP}
in a class of Hamiltonian perturbations of the quasi-linear
transport equation
\[
u_t+a(u)u_x=0,
\]
where $u=u(x,t)$, $x,u\in\mathbb{R}$, $t\in\mathbb R^+$,  and where
$a$ is an arbitrary regular function of $u$. We will restrict
ourselves to the case of the KdV hierarchy,
 which,  for any  $m\in\mathbb N$,  is a Hamiltonian perturbation of the equation
\begin{equation}
\label{mBurgers} u_{t_m} + C_m u^{m}u_x = 0 \qquad C_m =(-1)^{m+1}
\frac{2^m (2m+1)!!}{m!}.
\end{equation}
The equations in the KdV  hierarchy can be written in the form
\begin{equation}
\label{KdVm} u_{t_m}-(-1)^{m}\partial_x\psi_m(u,\e u_x,\e^2
u_{xx},\ldots,\e^{2m}\partial_x^{2m}u)=0,
\;\;t_m\in\mathbb{R}^+,\;\; m\in\mathbb N,
\end{equation}
where $\e>0$ and $\partial_x=\frac{\partial}{\partial x}$. The
function $\psi_m$ is polynomial in its variables and  it is the
variational derivative of the Hamiltonian $\mathcal{H}_m$
\cite{Gardner, GGKM, ZF}
\[
\psi_m(u,\e u_x,\ldots )=\dfrac{\delta \mathcal H_m}{\delta u(x)},\qquad \mathcal{H}_m=\int h_m(u, \e u_x(x), \e^2u_{xx}(x),\ldots)dx,
\]
which is defined as
\[
\dfrac{\delta \mathcal{H}_m}{\delta u(x)}=\dfrac{\partial h_m(u,\e
u_x,\ldots )}{\partial u}-\dfrac{d}{dx}\dfrac{\partial h_m(u,\e
u_x,\ldots)}{\partial u_x}+\dfrac{d^2}{dx^2}\dfrac{\partial h_m(u,\e
u_x,\ldots)}{\partial u_{xx}}-\ldots.
\]
The Hamiltonians  $\mathcal H_m$  satisfy the Lenard-Magri
recurrence relation \cite{Magri}
\begin{equation}
\begin{split}
\label{Lenard}
&\partial_x\dfrac{\delta \mathcal H_m}{\delta u(x)}=P\frac{\delta \mathcal
H_{m-1}}{\delta u(x)}, \;\;m\geq 1,\;\;\qquad \mathcal H_0[u]=\int
\dfrac{u^2(x)}{2}dx, \\
&P=\e^2\partial_x^3+4u\partial_x+2u_x.
\end{split}
\end{equation}
The densities $h_m$ are uniquely determined by (\ref{Lenard}) together with the conditions
\[
h_m(u,\e u_x,\e^2 u_{xx},\dots)|_{\e=0}=\dfrac{(-1)^{m+1}C_m}{(m+1)(m+2)}u^{m+2}.
\]
For $m=1$ the equation (\ref{KdVm})  coincides with the standard   KdV equation
\begin{equation}
\label{KdV} u_{t_1} + 6 u u_x + \epsilon^2 u_{xxx}  =  0,\qquad \e>0,
\end{equation}
 and for $m=2,3$ one has
the equations
\begin{align}
\label{KdV2} &u_{t_2} - 30 u^2 u_x - \epsilon^2 \left(20u_x u_{xx} +
10
u u_{xxx}\right) - \epsilon^4 u_{xxxxx} = 0,\\
 &u_{t_3}
+140u^3u_x+\e^2\left(70u_x^3+280uu_xu_{xx}+70u^2u_{3x}\right)\nonumber\\
&\qquad\qquad\qquad\qquad\qquad\quad+\e^4\left(70
u_{2x}u_{3x}+42u_xu_{4x}+ 14uu_{5x}\right)+\e^6u_{7x}=0\label{KdV3}.
\end{align}
The corresponding Hamiltonians for $m=1,2$ are \cite{GGKM1, GGKM2,
NMPZ}
\[
H_1=\int\left(u^3-\e^2\dfrac{u_{x}^2}{2}\right)dx,\quad H_2=\int\left(\dfrac{5}{2}u^4-5\e^2uu_{x}^2+\dfrac{\e^4}{2}u_{xx}^2\right)dx.
\]

Each equation in the hierarchy can also be written in the Lax form \cite{Lax}
\begin{equation}\label{LaxKdV}
L_{t_m}=[L,A_m],
\end{equation}
where $L$ is the Schr\"odinger operator
\begin{equation}\label{L}
L=\e^2\partial_x^2+u,
\end{equation}
$L_{t_m}$ is the operator of multiplication by $u_{t_m}$, and $A_m$
is an antisymmetric higher order operator with leading order
$(-1)^{m+1}4^{m}\e^{2m}\partial_x^{2m+1}$. The lower order terms are
determined by the requirement that $[L,A_m]$ is an operator of
multiplication with a function depending on $u, u_x, \ldots,
u_{(2m+1)x}$. For $m=1$ and $m=2$, we have
\begin{align}\label{P}
&A_1=4\epsilon^2\partial_x^3+3\left(u\partial_x+\partial_xu\right),\\
&A_2=-16\e^4\partial_x^5-20\e^2(\partial_x^3u+u\partial_x^3)+5(\partial_xu_{xx}+u_{xx}\partial_x)-15(\partial_xu^2+u^2\partial_x).
\end{align}
We will study the behavior of solutions to the KdV hierarchy in the
small dispersion limit where $\e\to 0$.

\medskip

 When
$\epsilon=0$ the KdV hierarchy reduces to  the transport equation
(\ref{mBurgers}). Let us assume that we have sufficiently smooth
negative initial data $u_0(x)$ with a single local minimum, and
which tend to $0$ rapidly at $\pm\infty$. The Cauchy problem for
(\ref{mBurgers}) can then be solved implicitly using the method of
characteristics, which leads to
\begin{equation}\label{char}
u(x,t_m)=u_0(\xi), \qquad -x+C_mu_0(\xi)^mt_m+\xi=0.
\end{equation}
This describes a left-moving solution for any $m\in\mathbb N$, but
the part of the solution near the minimum moves faster than the less
negative parts, so that the slope steepens at the left of the
negative hump as $t_m$ increases. The time $t_m^c$ where the slope
becomes vertical is called the time of gradient catastrophe, and is
given by
\begin{equation}\label{tc}
t_m^c=\frac{1}{m\max_{\xi\in\mathbb
R}\{-C_mu_0^{m-1}(\xi)u_0'(\xi)\}}.
\end{equation}
After this time, the solution to (\ref{char}) is no longer
single-valued.

In order to determine  the point of gradient catastrophe
$x^c$ and $u^c=u(x^c,t_m^c)$, one has to solve the system of three
equations
\begin{align}
\label{xc1}
&F(u;x,t_m):=-x+C_mu^mt_m+f_L(u)=0,\\
\label{tc1}
&F'(u;x,t_m)=mC_mu^{m-1}t_m+f_L'(u)=0,\\
\label{uc1} &F''(u;x,t_m)=m(m-1)C_mu^{m-2}t_m+f_L''(u)=0,
\end{align} for the unknowns $u, x, t_m$.
Here primes denote derivatives with respect to $u$, and $f_L$ is
the inverse of the decreasing part of the initial data $u_0$. Among
the possibly many solutions of (\ref{xc1})-(\ref{uc1}), the point of
gradient catastrophe is the solution $(u^c; x^c, t_m^c)$ with
minimal time $t_m^c$. We say that the gradient catastrophe is
generic if
\begin{equation}\label{k}
k:=-F'''(u^c;x^c,t_m^c)=-m(m-1)(m-2)C_m(u^c)^{m-3}t_m^c-f_L'''(u^c)\neq
0.
\end{equation}

\medskip

In the case $m=1$, it is well-known that the dispersive term
$\e^2u_{xxx}$ in (\ref{KdV}) regularizes the gradient catastrophe
that occurs for the Hopf equation $u_{t_1}+6uu_x=0$: the solution to
the KdV equation exists for all $t_1>0$ under suitable conditions on
the initial data $u_0(x)$. For $t_1<t_1^c$, the KdV solution is
approximated by the Hopf solution for small $\e>0$, and for $t_1>t_1^c$,
an interval of rapid oscillations is formed where the KdV solution
can be modeled using Jacobi elliptic $\theta$-functions \cite{GP,
LL, V2, DVZ, DVZ2, GK}.

For $m>1$ we have not been able to find results about  global
existence in time in the literature. For initial data in weighted
Sobolev spaces only local results stating that the solution exists
(and stays in the same space) for small times $t_m$ seem to be
available, see e.g.\ \cite{KPV}. For the class of analytic initial
data we will consider,  global existence in time should not be an
issue
 for any $m\in\mathbb N$.
 We will comment on this later,
 see Remark \ref{remark solution}.
However our aim is not to prove an existence result for the KdV hierarchy,
rather we want to study the asymptotic behavior for small $\e$ of
the solution $u(x,t_m,\e)$ to the Cauchy problem for (\ref{KdVm}) under the assumption that it exists.
 Before gradient catastrophe,
one does not expect to see a significant difference between
solutions to (\ref{mBurgers}) and (\ref{KdVm}) for small $\e$. After
the gradient catastrophe, the KdV hierarchy solution will also
develop a region of oscillatory behavior for small $\e$.

\medskip

The purpose of this manuscript is twofold. First, for $t_m<t_m^c$, we
will prove that $u(x,t_m,\e)=u(x,t_m)+\bigO(\e^2)$ as $\e\to 0$,
where $u(x,t_m)$ is the solution to the dispersionless equation.
Secondly, for $t_m\approx t_m^c, x\approx x^c$, we will show that
the KdV hierarchy solution can be approximated for small $\e$ by a
special Painlev\'e transcendent $U=U(X,T)$, which solves the fourth
order ODE \cite{Kapaev, BMP, dubcr, KS}
\begin{equation}\label{PI20}
X=T\, U -\left[ \dfrac{1}{6}U^3  +\dfrac{1}{24}( U_{X}^2 + 2 U\,
U_{XX} ) +\frac1{240} U_{XXXX}\right].
\end{equation}
This ODE is the second member of the Painlev\`e I hierarchy, and we
refer to it as the ${\rm P_I^2}$ equation. The relevant solution is real and has
the asymptotic behavior
     \begin{equation}
\label{PI2asym}
        U(X,T)=\mp (6|X|)^{1/3}\mp \frac{1}{3}6^{2/3}T|X|^{-1/3}
            +\bigO(|X|^{-1}),
            \qquad\mbox{as $X\to\pm\infty$,}
\end{equation}
for any fixed $T\in\mathbb{R}$, and has no poles for real values of
$X$ and $T$ \cite{CV1, Menikoff, Moore}. It is also remarkable that
$U(X,T)$ is an exact solution to the KdV equation normalized as
\begin{equation}\label{UKdV}U_T+UU_X+\frac{1}{12}U_{XXX}=0.\end{equation}

It was conjectured by Dubrovin in \cite{dubcr} that, for any
Hamiltonian perturbation of a hyperbolic equation
\cite{dubcr,Lorenzoni}, a generic solution $u(x,t,\e)$ has an
asymptotic expansion of the form
\begin{equation}
\label{univer} u(x,t,\e)= u^c +a_1\e^{2/7}U \left( a_2\e^{-6/7}(x-
x^c-a_3(t-t^c)), a_4\e^{-4/7}(t-t^c)\right) +o\left(
\epsilon^{2/7}\right),
\end{equation}
for $x, t$ near the point of gradient catastrophe $x^c, t^c$ of the
unperturbed equation, and with constants $a_1, a_2, a_3, a_4$
depending only on the initial data and on the equation. The
expansion should hold in a double scaling limit where $\e\to 0$, but
at the same time $x$ and $t$ should tend to the point $x^c$ and time
$t^c$ of gradient catastrophe for the unperturbed equation in such a
way that the arguments of $U$ in (\ref{univer}) remain bounded. In
other words, the limit is such that $\e\to 0$ and at the same time
$\e^{-6/7}(x- x^c-a_3(t-t^c))$ and $\e^{-4/7}(t-t^c)$ remain
bounded. The Painlev\'e transcendent $U(X,T)$ is thus conjectured to
describe the behavior of the solution to the perturbed equation near
the point of gradient catastrophe for the unperturbed equation, and
is expected to be universal in the sense that it is independent of
the choice of the equation and independent of the choice of initial
data. The only quantities in (\ref{univer}) that depend on the
initial data and on the equation are the constants $a_1, a_2, a_3,
a_4$, and the values of $x^c, t^c, u^c$. The asymptotic formula
(\ref{univer}) was shown numerically for a certain class of
equations including the KdV equation and the second member of the
KdV hierarchy \cite{GK1, DGK1}.

We prove this conjecture in the special case of the KdV hierarchy
for a class of analytic initial data with a single negative hump. So
far, the conjecture had been proven only for the KdV equation
\cite{CG}. Similar results appear also in double scaling limits for Hermitian random matrix ensembles \cite{CV2}  and in the semiclassical limit of the focusing nonlinear Schr\"odinger equation \cite{DGK}.

\subsection{Statement of results}

We study the Cauchy problem for equation (\ref{KdVm}) with
$m\in\mathbb N$. Similarly as in \cite{CG}, we impose the following
conditions on the initial data $u_0$.
\begin{Assumptions}\label{assumptions}\
\begin{itemize}
\item[(a)] $u_0(x)$ is real analytic and has an analytic continuation to the complex plane in the domain
\[
\mathcal{S}=\{z\in\mathbb{C}: |\Im z|<\tan\theta|\Re
z|\}\cup\{z\in\mathbb C: |\Im z|<\sigma\}
\]
where $0<\theta<\pi/2$ and $\sigma>0$;
\item[(b)] $u_0(x)$ decays as $|x|\rightarrow\infty$ in $\mathcal{S}$ such that
\begin{equation}
u_0(x)=\bigO\left(\frac{1}{|x|^{3+s}}\right), \;\;s>0, \quad x\in
\mathcal S,
\end{equation}
\item[(c)] for real $x$,  $u_0(x)<0$ and $u_0$ has a single local minimum at a certain point
$x_M$, with
\[u_0'(x_M)=0, \qquad u_0''(x_M)>0.\] Without loss of generality, we assume that $u_0$ is normalized such
that $u_0(x_M)=-1$.
\end{itemize}
\end{Assumptions}

\medskip

We prove the following result.

\begin{theorem}\label{theorem: 1}
Let $u_0(x)$ satisfy the conditions described in Assumptions
\ref{assumptions}, and let $m\in\mathbb N$, $x\in\mathbb R$ and
$t_m<t_m^c$, where $t_m^c$ is the time of gradient catastrophe for
(\ref{mBurgers}) given by (\ref{tc}).   If  $u(x, t_m,\e)$
solves equation (\ref{KdVm}) with initial condition $u(x,0,\e)=u_0(x)$,  then we have
\begin{equation}\label{u0}
u(x,t_m,\e)=u(x,t_m)+\bigO(\e^2), \qquad \mbox{ as $\e\to 0$,}
\end{equation}
where $u(x,t_m)$ is the solution to the Cauchy problem for
(\ref{mBurgers}), given by (\ref{char}).
\end{theorem}

\begin{remark}
We will prove this result in detail for values of $x,t_m$ where
$u(x,t_m)$ is decreasing, i.e.\ for $x<x_M+C_mt_m$ since $x_M+C_mt_m$ is the position of the minimum at time $t_m$. For $x\geq
x_M+C_mt_m$, the proof is similar, but several technical
modifications are needed. We will discuss those changes in Remark
\ref{remark: increasing}.
\end{remark}

Our second result describes the behavior of $u(x,t_m,\e)$ when
$x,t_m$ approach the point and time of gradient catastrophe at
appropriate speeds. We will prove that (\ref{univer}) holds, with
the values of the constants given by
\begin{align}&\label{a12}
a_1=\frac{2}{(8k)^{2/7}} ,&&a_2=\frac{1}{(8k)^{1/7}},\\
\label{a34}&a_3=C_m(u^c)^m,&&
a_4=\frac{2mC_m(u^c)^{m-1}}{(8k)^{3/7}}.\end{align} We will give an
asymptotic expansion as $\e\to 0$ for $(x,t_m)$ in a space-time
window of size $\bigO(\e^{4/7})$, and in addition $x-x_c$ must be
equal to $C_m(u^c)^m(t-t_m^c)$ plus a correction of size
$\bigO(\e^{6/7})$. In this shrinking (as $\e\to 0$) region in the
$(x,t_m)$-plane, the transition takes place between asymptotics
determined by (\ref{mBurgers}) and the oscillatory asymptotics that
are expected to be present for $t_m>t_m^c$. The transition is
described by the Painlev\'e transcendent $U(X,T)$. In addition, we
will also compute the next term in the asymptotic expansion
(\ref{univer}), which is of order $\bigO(\e^{4/7})$. This term is
rather complicated but can still be expressed completely in terms of
the Painlev\'e transcendent $U(X,T)$. We have an expansion of the
form
\begin{multline}\label{expansionu}
u(x,t_m,\e) =u_c+a_1\e^{2/7} U
+c_1\epsilon^{4/7}\left( QU_X+U_{XX}+4U^2-3c_2U_T\right)\\
+c_3\epsilon^{4/7} (2U_XQ_T+4UU_T+\dfrac{1}{2}U_{XXT})+
\bigO(\epsilon^{5/7}),
\end{multline}
where we used the abbreviations
\begin{align}
&U=U\left( a_2\e^{-6/7}(x- x^c-a_3(t-t^c)),
a_4\e^{-4/7}(t-t^c)\right),\\
&Q=\dfrac{1}{240}U_XU_{XXX}-\dfrac{U_{XX}^2}{480}+XU-\dfrac{T}{2}U^2+\dfrac{U^4}{24}+\dfrac{1}{24}UU_X^2.
\end{align}
The values of $c_1, c_2, c_3$ are
\begin{align}
&\label{c1}c_1=\dfrac{32F^{(4)}(u^c)}{63(8k)^{11/7}},\\
&c_2=\frac{(x-x^c)-C_m(u^c)^m(t_m-t_m^c)}{(8k)^{1/7}\e^{6/7}},\\
&\label{c3}c_3=\dfrac{mC_m(u^c)^{m-1}(t_m-t^c_m)}{4k\e^{4/7}}\left(\dfrac{2(m-1)}{5u^c}+\dfrac{2F^{(4)}(u^c)}{21k}\right),
\end{align}
with $k$ given by (\ref{k}), and $F$ by (\ref{xc1}).

\begin{theorem}\label{theorem: 2}
Let $u_0(x)$ satisfy the conditions described in Assumptions
\ref{assumptions} and assume that the generic condition
\begin{equation}\label{knot0}
k:=-F'''(u^c)=-m(m-1)(m-2)(u^c)^{m-3}-f_L'''(u^c)\neq 0
\end{equation}
holds with $m\in\mathbb N$. Write $u^c, x^c, t_m^c$ for the solution
to the system (\ref{xc1})-(\ref{tc1})-(\ref{uc1}).
Let  us take a double scaling limit where  $\epsilon\to 0$ and
at the same time  $x\to x^c$ and $t_m\to t^c_m$ in such a way
that, for some $X,T\in\mathbb R$,
\begin{equation}\label{double scaling}
\lim\dfrac{x-
x^c-C_m(u^c)^m(t_m-t^c_m)}{(8k)^{\frac{1}{7}}\e^{\frac{6}{7}}}=X,\quad
\lim\dfrac{2mC_m(u^c)^{m-1}(t_m-t^c_m)}{(8k)^{\frac{3}{7}}\epsilon^{\frac{4}{7}}}=T.
\end{equation}
If   $u(x,
t_m,\e)$ solves equation (\ref{KdVm})  with initial
condition $u(x,0,\e)=u_0(x)$,   the asymptotic expansion
(\ref{expansionu}) holds in the double scaling limit.
\end{theorem}

The proofs of our results are based on the direct and inverse
scattering transform for the KdV hierarchy. This approach relies on
the Lax representation (\ref{LaxKdV}), and the inverse scattering
transform can be formulated as a Riemann-Hilbert problem, where one
searches for a function which satisfies a prescribed jump condition,
depending on the reflection coefficient for the Schr\"odinger
equation $Lf=\lb f$. Solving the RH problem asymptotically as $\e\to
0$ leads to small dispersion asymptotics for $u(x,t_m,\e)$. We will
use a Deift/Zhou steepest descent method similar to the one in
\cite{DVZ, DVZ2} for the asymptotic analysis of the RH problem. This
method consists of a series of transformations $M\mapsto T\mapsto
S\mapsto R$ of the RH problem, which results at the end in a RH
problem for $R$ which can be solved approximately for small $\e$.
The first transformation $M\mapsto T$ involves the construction of a $G$-function
satisfying convenient jump and asymptotic conditions. The second
transformation $T\mapsto S$ deforms the jump contour from the real
line to a lens-shaped contour. The last transformation $S\mapsto R$
requires the construction of local and global parametrices. A local
Airy parametrix will be needed for the proof of Theorem
\ref{theorem: 1}. For the proof of Theorem \ref{theorem: 2}, we will
need to build a local parametrix out of a model RH problem related
to the ${\rm P_I^2}$ equation. The most important new features here compared to
\cite{DVZ, CG} are the generalization of the $G$-function to the
case $m>1$, and the generalized construction of the local ${\rm P_I^2}$
parametrix.

 \section{Riemann-Hilbert problem for the KdV hierarchy\label{section: inverse scattering}}
We construct a RH problem using particular solutions to the
Schr\"odinger equation $Lf=\lambda f$, with $L$ given by (\ref{L})
with potential $u=u(x,t_m,\e)$. This construction is well understood
\cite{DeiftTrubowitz, BDT, Shabat, CG, Faddeev}, but we summarize the main
lines here for the convenience of the reader.

For negative $u$, the Schr\"odinger operator has no point spectrum.
Moreover, if $u$ solves the equation (\ref{KdVm}), as a consequence
of the Lax equation (\ref{LaxKdV}), the eigenvalues of $L$ are
independent of $t_m$. Since our initial data $u_0$ are negative, it
follows that $L$ has no point spectrum at any time $t_m>0$. If
\begin{equation}\int_{-\infty}^{+\infty}|u(x,t_m,\e)|(1+x^2)dx<\infty,\end{equation}
there exist \cite{DeiftTrubowitz} fundamental Jost solutions
$\psi_\pm=\psi_\pm(\lb;x,t_m,\e)$ and
$\phi_\pm=\phi_\pm(\lb;x,t_m,\e)$ to the Schr\"odinger equation
satisfying the asymptotic conditions
\begin{align}
\label{normalization1}
&\lim_{x\to +\infty}\psi_\pm(z;x,t_m,\epsilon) e^{\pm\frac{i}{\e}\sqrt{-\lb} x}=1,\\
&\label{normalization2}\lim_{x\to -\infty}\phi_\pm(z;x,t_m,\epsilon)
^{\mp\frac{i}{\e}\sqrt{-\lb}x}=1,
\end{align}
for $\lambda\in\mathbb C\setminus\{0\}$. We fix $\sqrt{-\lambda}$ to
be the principal branch of the square root which is analytic for
$\lambda\in\mathbb C\setminus\mathbb R^+$ and positive for
$\lambda<0$. We consider $\psi_\pm$ and $\phi_\pm$ as functions in
the variable $\lb$, whereas $x,t_m,\e$ will be parameters. Those
solutions can be constructed as a solution to Volterra integral
equations as in \cite{DeiftTrubowitz}, and the analysis of the
integral equations shows that $\psi_-$ and $\phi_-$ can be continued
analytically for $\lambda$ in the lower half plane, and $\psi_+$ and
$\phi_+$ to the upper half plane. Also asymptotics as $\lb\to\infty$
for $\phi_\pm$ and $\psi_\pm$ can be deduced.

The fundamental solutions $\psi_\pm$ and $\phi_\pm$ are related as follows,
\begin{equation}
\label{transition}
\begin{pmatrix}\psi_+(\lb)&\psi_-(\lb)\end{pmatrix}=\begin{pmatrix}\phi_-(\lb)&\phi_+(\lb)\end{pmatrix}
\begin{pmatrix} a(\lb;t_m,\e)& \overline b(\lb;t_m,\e)\\
 b(\lb;t_m,\e)& \overline a(\lb;t_m,\e)
\end{pmatrix},\qquad \lb<0,
\end{equation}
with
\begin{equation}
\label{normalizationab} |a(\lb)|^2-|b(\lb)|^2=1,\qquad \mbox{for
}\lb<0,
\end{equation}
and where $a$ and $b$ are independent of $x$. The quantities
\[
r(\lb;t_m,\e):=\frac{b(\lb;t_m,\e)}{a(\lb;t_m,\e)},\qquad
t(\lb;t_m,\e):=\dfrac{1}{a(\lb;t_m,\e)},
\]
are the reflection and transmission coefficients (from the left) for
the Schr\"odinger equation and depend on $t_m,\e$ through
$u(x,t_m,\e)$. They are continuous for $\lambda\leq 0$. In
particular we have
\begin{equation}
\label{transition1} \frac{\psi_+}{a}\sim\begin{cases}
e^{-\frac{i}{\e}\sqrt{-\lb}x}+\frac{b}{a}e^{\frac{i}{\e}\sqrt{-\lb}x},& x\to -\infty,\\
\frac{1}{a}e^{-\frac{i}{\e}\sqrt{-\lb}x},&x\to +\infty.
\end{cases}
\end{equation}
If $u$ solves the higher order KdV equation (\ref{KdVm}), the
Gardner-Greene-Kruskal-Miura \cite{GGKM} relations are
\begin{equation}\label{abt}
\dfrac{da}{dt_m}=0,\quad \dfrac{db}{dt_m}=\frac{2i}{\e}
4^mt_m(-\lb)^{\frac{2m+1}{2}}b.
\end{equation}
Indeed, (\ref{LaxKdV}) implies that $\frac{d}{dt}\psi_++A_m \psi_+$
is also a solution to the Schr\"odinger equation. The asymptotics as
$x\to +\infty$ then imply that
\[
\frac{d}{dt_m}\psi_+=-A_m\psi_++\dfrac{4^mi}{\e}(-\lb)^{\frac{2m+1}{2}}\psi_+.
\]
Together with (\ref{transition1}) this implies (\ref{abt}) and thus
\[
r(\lb;t_m,\e)=r(\lb;0,\e)e^{\frac{8i}{\e}t_m(-\lb)^{\frac{2m+1}{2}}}.
\]
The transmission coefficient is analytic for $\lb$ in the upper half
plane, and the reflection coefficient is analytic in a region of the
form $\{\pi -\theta_0 < \arg \lb < \pi\}$, with $\theta_0>0$. If the
potential $u$ is smooth in $x$, the reflection coefficient decays
rapidly as $\lb\to - \infty$.

We will now construct the solution to a RH problem using the Jost
solutions $\psi_\pm$ and $\phi_\pm$: write $M=M(\lambda;x,t_m,\e)$
by
\begin{equation}\label{M}
M(\lb;x,t_m,\e)=\begin{cases}\begin{pmatrix}\phi_+&\frac{1}{a}
\psi_+\\
\e\frac{d}{dx}\phi_+&\frac{\e}{a}\frac{d}{dx}\psi_+\end{pmatrix}e^{\frac{-i}{\e}\alpha(\lambda;x,t_m)\sigma_3},&
\mbox{ as $\lambda\in\mathbb C^+$},\\[3ex]
\begin{pmatrix}\frac{1}{a^*}\psi_-&\phi_-\\
\frac{\e}{a^*}\frac{d}{dx}\psi_-&\e\frac{d}{dx}\phi_-\end{pmatrix}e^{\frac{-i}{\e}\alpha(\lambda;x,t_m)\sigma_3},&\mbox{
as $\lb\in\mathbb C^-$},
\end{cases}
\end{equation}
where
\begin{align}
&\phi_\pm=\phi_\pm(\lambda;x,t_m,\e),\quad
\phi_\pm=\phi_\pm(\lb;x,t_m,\e), \\
&a=a(\lb;\e),\quad
a^*=\overline{a(\overline\lambda;\e)},
\end{align}
and
\begin{equation}\label{alpha}
\alpha(\lambda;
x,t_m)=x(-\lb)^{\frac{1}{2}}+4^mt_m(-\lambda)^{m+\frac{1}{2}}.
\end{equation}
Using (\ref{transition}) and the asymptotics for the
Jost solutions as $\lambda\to\infty$, one shows that $M$ solves a RH
problem, see \cite{BDT, Shabat, CG}:

\subsubsection*{RH problem for $M$}
\begin{itemize}
\item[(a)] $M(\lb;x,t_m,\e)$ is analytic for
$\lb\in\mathbb{C}\backslash \mathbb{R}$, \item[(b)] $M$ has
continuous boundary conditions $M_\pm(\lambda)$ for
$\lambda\in\mathbb R\setminus\{0\}$ that satisfy the jump conditions
\begin{align}&\label{RHP M1}M_+(\lb)=M_-(\lb){\small \begin{pmatrix}1&
r_0(\lb;\e)e^{2i\alpha(\lb;x,t_m)/e}\\-\bar{r_0}(\lb;\e)
e^{-2i\alpha(\lambda;x,t_m)/\e}&1-|r_0(\lb;\e)|^2
\end{pmatrix}}&\mbox{ for $\lb<0$,}\\
&M_+(\lb)=M_-(\lb)\sigma_1,\quad
\sigma_1=\begin{pmatrix}0&1\\1&0\end{pmatrix}&\mbox{ for $\lb>0$},
\end{align}
with $r_0(\lambda;\e):=r(\lb;0,\e)$.
\item[(c)] We have
\begin{equation}\label{RHP M:c}M(\lb;x,t_m,\e)=
\begin{pmatrix}1&1\\&\\i\sqrt{-\lb}&-i\sqrt{-\lb}\end{pmatrix}\left(I-\frac{q}{2i\e\sqrt{-\lambda}}\sigma_3+\bigO(\lambda^{-1})\right),
\mbox{ for $\lb\rightarrow \infty$},
\end{equation}
with $q=q(x,t_m,\e)$ independent of $\lambda$, and $u=q_x$ is the
solution to equation (\ref{KdVm}) with initial data $u_0$.
\end{itemize}

In other words, the solution to the KdV hierarchy with initial data
$u_0$ can be recovered from the solution to the RH problem by the
formula
\begin{equation}\label{uM}
u(x,t_m,\e)=-2i\e\partial_x M_{1, 11}(x,t_m,\e),\qquad
\partial_x=\frac{\partial}{\partial x},
\end{equation}
where $M_{11}(\lb;x,t_m,\e)=1+\dfrac{M_{1,
11}(x,t_m,\e)}{\sqrt{-\lb}}+\bigO(1/\lb)$ as $\lb\to\infty$.

\begin{remark}\label{remark solution}
Using the vanishing lemma approach developed in \cite{FokasZhou,
FokasMuganZhou, FIKN}, one can show that the RH problem for $M$ is
solvable for any value of $x,t\in\mathbb R$, $\e>0$ if $r_0$ has
sufficient regularity and sufficient decay at $-\infty$ and if
$|r_0(\lambda)|<1$ for $\lambda<0$. The solvability of the RH
problem can be used to prove that the Cauchy problem for equation
(\ref{KdVm}) is solvable for initial data in a suitable space
following the proofs in \cite{Zhou}. \end{remark}

\section{Asymptotic analysis of the RH problem as $\e\to 0$}

This section contains the asymptotic analysis of the RH problem for
$M$ as $\e\to 0$. We will study the RH problem for fixed $x,t_m$
(for the proof of Theorem \ref{theorem: 1}), and in a double scaling
limit for $x\approx x^c$, $t_m\approx t_m^c$ (for the proof of
Theorem \ref{theorem: 2}). The construction of the $G$-function in
Section \ref{section: g} is almost the same in both cases, and also
for the transformations $M\mapsto T\mapsto S$ in Sections
\ref{section: T} and \ref{section: S} and the construction of the
outside parametrix in Section \ref{section: outside}, we do not need
to distinguish between the regular and the critical case. It is only
when we construct a local parametrix that there is an essential
difference. For the regular case, we assume that $x$ and $t_m$ are
such that $u_x(x,t_m)<0$, i.e.\ we consider only the decreasing part
where $x<x_M+C_mt_m$, the position of the minimum at time $t_m$. For the increasing part, several changes have
to be made, see Remark \ref{remark: increasing}.

\subsection{The $G$-function}\label{section: g} In this section,
we will define a $G$-function, which will be needed to modify the
jumps of the RH problem in a suitable way. Let us first write the
Abel transform
\[
F_A(\lambda;x,t_m)=\dfrac{1}{2}\int^0_{\lambda}\frac{F(\xi;x,t_m)d\xi}{\sqrt{\xi-\lambda}},\qquad\mbox{ for $\lambda\in
[-1,0)$,}
\]
where $F$ is defined by (\ref{xc1}). We define the $G$-function
$G=G(\lambda;x,t_m; u)$ as
\begin{equation}
\label{def Godd} G(\lambda;x,t_m;u)=\frac{\sqrt{u -
\lambda}}{\pi}\int_{u}^0\frac{F_A(\eta;x,t_m)}{(\eta
-\lambda)\sqrt{\eta-u}}d\eta.
\end{equation}

We will have to make two different choices for $u\in (-1,0)$. For
the proof of Theorem \ref{theorem: 1}, we need to choose
$u=u(x,t_m)$ to be the solution to equation (\ref{mBurgers}): this
solution is given by (\ref{char}) or, equivalently, by (\ref{xc1}).
For the proof of Theorem \ref{theorem: 2} on the other hand, we fix
$u$ to be $u=u^c=u(x^c,t_m^c)$. Whenever the choice of $u$, $x$, and
$t_m$ is unimportant below, we will simply write $G_m(\lambda)$ for
$G_m(\lambda;x,t_m;u)$.

\medskip

\begin{figure}[!htb]
\centering
\includegraphics[width=.6\textwidth]{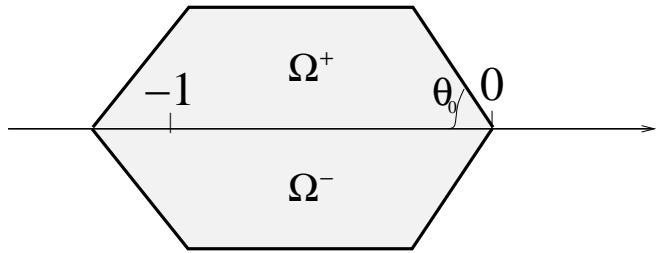}
\caption{The regions $\Omega^{+}$ and $\Omega^-$.}
\label{OMEGA}
\end{figure}

$G$ is analytic for $\lambda\in\mathbb C\setminus [u,+\infty)$. As
$\lambda\to \infty$, we have $G(\lambda)=\bigO(\lambda^{-1/2})$,
and writing
\begin{eqnarray}\label{g1}
{G}_{1}(x,t_m;u):=\lim_{\lambda\to
\infty}(-\lambda)^{1/2}G(\lambda;x,t_m,u),
\end{eqnarray}
one verifies that, for both choices of $u$ made above, we have the
identity
\begin{equation}
\partial_x {G}_1(x,t_m)=\frac{u}{2}.\label{dg1}
\end{equation}
For $\lambda$ on the branch cut, $G$ satisfies the jump properties
\begin{align}
&\label{prop g1}G_{+}(\lambda)+G_{-}(\lambda)=0,&\mbox{ for $\lambda\in(0,+\infty)$,}\\
&\label{prop g} G_{+}(\lambda)+
G_{-}(\lb)-2\rho(\lambda)+2\alpha(\lambda)=0,&\mbox{ for
$\lambda\in(u,0)$},
\end{align}
with $\rho$  given by
\begin{equation}\label{def rho}
\rho(\lb)=\frac{1}{2}
\int_\lambda^0\frac{f_L(\xi)}{\sqrt{\xi-\lambda}}d\xi.
\end{equation}
The function $\rho(\lambda)$ is clearly well-defined for
$-1<\lambda<0$, but we can extend it to an analytic function.
Because of the analyticity of the initial data $u_0$, $\rho$ is
analytic in a neighborhood of $(-1,0)$. It cannot be extended to an
analytic function in a full neighborhood of $-1$ because the inverse
$f_L$ behaves like a square root near $-1$, but we can extend it to
an analytic function in a neighborhood of $[-1,0)$, except for a
branch cut which we choose along $\lb<-1$. We write
$\Omega=\Omega^+\cup\Omega^-\cup(-1,0)$ for such a region
where $\rho$ is analytic, with $\Omega^+\subset \mathbb C^+$ and $\Omega^-=\overline{\Omega^+}$. For
later convenience, we choose $\Omega^+$ sufficiently small so that
it is contained in the region $\{\pi-\theta_0<\arg\lambda < \pi\}$
where $r_0(\lambda)$ is analytic (see Fig.~\ref{OMEGA})

Let us define an auxiliary function $\phi$ in $\Omega^+\cup\Omega^-$ by
\begin{equation}
\label{definition
phi}\phi(\lb;x,t_m)=G(\lb;x,t_m)-\rho(\lb)+\alpha(\lb;x,t_m),
\end{equation}
so that $\phi$ is analytic across $(-1,u)$, but not on
$(-1-\delta,-1)$, because this is a part of the branch cut for
$\rho$, and not on $(u,0)$, because this is part of the branch cut
for $G$. It is an analytic function for
$\lambda\in\Omega\setminus[u,0)$. By (\ref{prop g}), we have
\begin{equation}2\phi_{+}(\lb;x,t_m)=G_+(\lb;x,t_m)-
G_-(\lb;x,t_m),\qquad \text {for $\lb\in (u,0)$}.\label{def phi
2}\end{equation} After a straightforward integral calculation as in
\cite{CG}, one observes that $\phi$ can be written as
\begin{eqnarray}
\phi(\lambda;x,t_m)&=&\dfrac{1}{2}\int_{u}^{\lambda}\dfrac{F(\xi;x,t_m)}{\sqrt{\xi-\lambda}}d\xi\label{phi0odd1}\\
&=&-\sqrt{u-\lambda}
F(u;x,t_m)+\dfrac{2}{3}(u-\lambda)^{\frac{3}{2}}F'(u;x,t_m)\nonumber\\
& &-\dfrac{4}{15}(u-\lambda)^{\frac{5}{2}}F''(u;x,t_m)
-\dfrac{4}{15}\int_{u}^{\lambda}F'''(\xi;x,t_m)(\xi-\lambda)^{\frac{5}{2}}d\xi.\label{phi0odd}
\end{eqnarray}

\subsection{First transformation $M\mapsto T$}\label{section:
T}
We are now ready to perform a first transformation of the RH
problem. This will lead to more convenient jump matrices for the
asymptotic analysis as $\e\to 0$. A crucial role is played by the
$G$-function and its properties discussed before.

\medskip

Define
\begin{equation}\label{def T}
T(\lambda;x,t_m,\e)=\begin{pmatrix}1&0\\
\frac{q}{2}&1\end{pmatrix}M(\lambda;x,t_m,\e)e^{-\frac{i}{\epsilon}G(\lambda;x,t_m)\sigma_3}.
\end{equation}
We then have

\subsubsection*{RH problem for $T$}
\begin{itemize}
\item[(a)] $T$ is analytic in $\mathbb C\setminus \mathbb R$,
\item[(b)] $T_+(\lambda)=T_-(\lambda)v_T(\lambda)$, \ \ \ as
$\lambda\in\mathbb R$,
with
\begin{equation}\label{vT}v_T(\lambda)=e^{\frac{i}{\epsilon}\mathcal
G_-(\lambda)\sigma_3}v_M(\lambda) e^{-\frac{i}{\epsilon}\mathcal
G_+(\lambda)\sigma_3},\end{equation} and $v_M$ is the jump matrix in the RH problem for $M$ given by
\begin{equation}
v_M(\lb)=\begin{cases}
\begin{pmatrix}1&
r_0(\lb;\e)e^{2i\alpha(\lb;x,t_m)/e}\\-\bar{r_0}(\lb;\e)
e^{-2i\alpha(\lambda;x,t_m)/\e}&1-|r_0(\lb;\e)|^2
\end{pmatrix},&\mbox{ for $\lb<0$,}\\
\sigma_1,&\mbox{ for $\lb>0$}.
\end{cases}
\end{equation}
\item[(c)]As $\lambda\to\infty$,
\begin{equation}\label{RHP T:c}T(\lambda)=\left(I+\bigO(\lambda^{-1})\right)\begin{pmatrix}1&1\\
i\sqrt{-\lambda}&-i\sqrt{-\lambda}\end{pmatrix}.
\end{equation}
\end{itemize}

The left multiplication with the triangular matrix in (\ref{def T})
was needed to transform (\ref{RHP M:c}) in (\ref{RHP T:c}), but has
no effect on the jumps for $T$. The jump matrix (\ref{vT}) can be
simplified using the properties (\ref{prop g1})-(\ref{prop g}) of
the $G$-function and the definition (\ref{definition phi}) of
$\phi$. We write the jump matrix in a different form depending on
the value of $\lambda\in\mathbb R$. For $\lambda>0$, by (\ref{prop
g1}) we have
\begin{equation}\label{vT1}
v_T(\lambda)=\sigma_1.
\end{equation}
For a sufficiently small choice of $\delta_1>0$, we can write the
jump matrix in terms of the previously defined function $\phi$ on
the interval  $(-1-\delta_1, 0)$. For $\lambda\in(u,0)$, we use
(\ref{prop g}) and (\ref{def phi 2}) to conclude that
\begin{equation}\label{vT2}
v_T(\lambda)=\begin{pmatrix}e^{-\frac{2i}{\e}\phi_+(\lb)}
&i\kappa_+(\lb)\\
i\kappa_-^*(\lb)&(1-|r_0(\lb)|^2)e^{\frac{2i}{\e}\phi_+(\lb)}
\end{pmatrix},\qquad\mbox{ as $\lambda\in (u,0)$,}
\end{equation}
where we have written $\kappa$ for
\begin{equation}\label{def kappa}
\kappa(\lambda;\e)=-ir_0(\lb;\e)e^{\frac{2i}{\e}\rho(\lb)},\qquad
\mbox{ for $\lambda\in\Omega^+$},
\end{equation}
with boundary values on $\mathbb R$ denoted by $\kappa_+(\lambda)$
and $\kappa_-^*(\lb)=\bar\kappa_+(\lambda)$. For
$\lambda\in(-1-\delta_1, u)$, by (\ref{prop g}) and (\ref{definition
phi}),
\begin{equation}
v_T(\lambda)=
\begin{pmatrix}1&i\kappa_+(\lb)e^{\frac{2i}{\e}\phi_+(\lb)}\\
i\kappa_-^*(\lb)e^{-\frac{2i}{\e}\phi_-(\lb)} &1-|r_0(\lb)|^2
\end{pmatrix}.
\end{equation} Here, the boundary values $\phi_\pm$
are needed only on $(-1-\delta_1, -1]$, on $(-1,u)$, we have
$\phi=\phi_\pm$. Finally, on $(-\infty,-1-\delta_1)$, we have
\begin{equation}
v_T(\lambda)=\begin{pmatrix}1&r_0(\lb)e^{\frac{2i}{\e}(\mathcal
G(\lb)+\alpha(\lb))}\\
-\bar r_0(\lb)e^{-\frac{2i}{\e}(\mathcal G(\lb)+\alpha(\lb))}
&1-|r_0(\lb)|^2
\end{pmatrix},\qquad\mbox{ as $\lambda<-1-\delta_1$}.
\end{equation}

 Using (\ref{dg1}),
(\ref{def T}), and (\ref{uM}), we recover the solution of the higher
order KdV equation by
\begin{eqnarray}
u(x,t_m,\e)&=&2\partial_x \mathcal{G}_1(x,t_m;u) -2i\e\partial_x T_{11}^1(x,t_m,\e)\nonumber \\
&=&u-2i\e\partial_x T_{11}^1(x,t_m,\e),\label{uT}
\end{eqnarray}
where $T_{11}^1$ is given by
\[T_{11}(\lb;x,t_m,\e)=1+
\frac{T_{11}^1(x,t_m,\e)}{\sqrt{-\lb}}+O(\lb^{-1}),\qquad \mbox{ as
$\lb\to\infty$}.\]

\medskip

The aim of this RH analysis is to end up with jump matrices that
decay to the identity matrix when $\e\to 0$. To get a feeling for
the small $\e$ behavior of the jump matrix $v_T$, we need to have
information about the reflection coefficient $r_0(\lambda;\e)$ for
small values of $\e$. We have the following results, see \cite{CG, CG2} in combination with \cite{Ramond}, for any choice of $\delta_1>0$.
\begin{itemize}
\item[(i)] For $\lambda<-1-\delta_1$,
\begin{equation}\label{r0}
r_0(\lambda;\e)=\bigO(e^{-c\sqrt{-\lambda}/\e}), \qquad \mbox{ as
$\e\to 0$},
\end{equation}
with $c>0$.
\item[(ii)] for $\lambda$ lying in a region $\Omega_+$ as defined in Section \ref{section: g}, but
$\lambda$ bounded away from $-1$, say
$|\lambda+1|>\frac{\delta_1}{2}$, we have
    \begin{equation}\label{kappa1}
    \kappa(\lambda;\e)=1+\bigO(\e), \qquad \mbox{ as $\e\to 0$},
    \end{equation}
\item[(iii)] for $\lambda\in(u+\delta_1, 0)$, we have
\begin{equation}\label{r2}
(1-|r_0(\lambda)|^2)e^{\frac{2i}{\e}\phi_+(\lb)}=\bigO(e^{-c/\e}),
\qquad c>0, \qquad \mbox{ as $\e\to 0$.}
\end{equation}
The latter was shown in \cite{CG} for $m=1$ and $u=u^c$ only, but
the same argument applies to the case $m>1$ and $u=u(x,t_m)$.
\end{itemize}

This implies that the jump matrix $v_T(\lambda)$ tends to
$i\sigma_1$ for $\lambda\in(u,0)$, and that it tends to $I$ for
$\lambda<-1-\delta_1$. On $(-1-\delta_1, u)$, the jump matrix is
oscillatory for small $\e$. In the next section, we will deform the
contour in such a way that the oscillatory behavior turns into
exponential decay.

\subsection{Opening of the lens $T\mapsto S$}\label{section: S}
\begin{figure}[t]
\begin{center}
    \setlength{\unitlength}{1.4mm}
    \begin{picture}(137.5,26)(22,11.5)
        \put(112,25){\thicklines\circle*{.8}}
        \put(45,25){\thicklines\circle*{.8}}
        \put(47,26){\small $-1-\delta_1$}
        \put(112,26){\small $0$}\put(122,27){\small $\sigma_1$}
        \put(85,25){\thicklines\circle*{.8}} \put(80,26){\small $u$}
        \put(99,25){\thicklines\vector(1,0){.0001}}
        \put(85,25){\line(1,0){45}}
        \put(124,25){\thicklines\vector(1,0){.0001}}
        \put(22,25){\line(1,0){23}}
        \put(35,25){\thicklines\vector(1,0){.0001}}
\put(74,13){\small
$\begin{pmatrix}1&0\\i\kappa^* e^{-\frac{2i}{\epsilon}\phi}&1\end{pmatrix}$}
        \qbezier(45,25)(65,45)(85,25) \put(66,35){\thicklines\vector(1,0){.0001}}
        \qbezier(45,25)(65,5)(85,25) \put(66,15){\thicklines\vector(1,0){.0001}}
\put(73,36){\small
$\begin{pmatrix}1&i\kappa e^{\frac{2i}{\epsilon}\phi}\\0&1\end{pmatrix}$}
\put(34,27){\small $v_T$} \put(90,29){\small $\begin{pmatrix}
e^{-\frac{2i}{\epsilon}\phi_+}&i\kappa\\i\bar\kappa& o(1)
\end{pmatrix}$}
\put(64.5,28){I} \put(64,20){II} \put(49,32){$\Sigma_1$}
\put(49,16){$\Sigma_2$} \multiput(45,25)(1,0){40}{\circle*{0.1}}
    \end{picture}
    \caption{The jump contour $\Sigma_S$ and the jumps for $S$}
    \label{figure: sigmaS}
\end{center}
\end{figure}
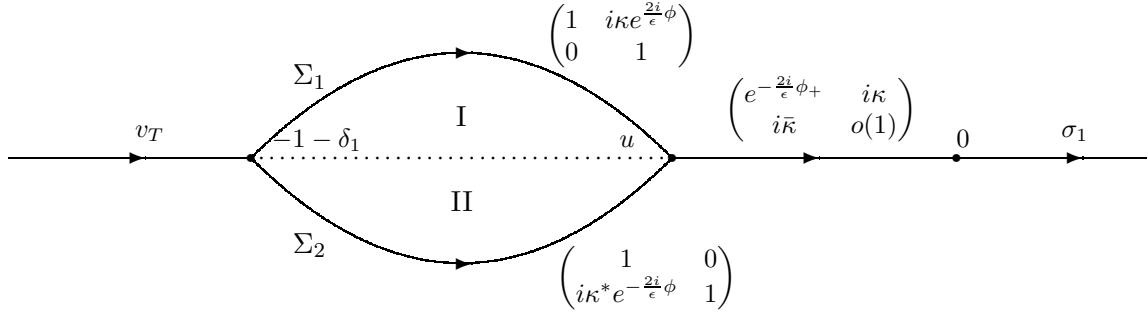

The jump matrix $v_T(\lambda)$ can be written in the following
factorized form for $-1-\delta_1<\lambda<u$,
\begin{equation}
v_T(\lb)=\begin{pmatrix}1&0\\
i{\kappa}^*_-(\lb)e^{-\frac{2i}{\e}
\phi_-(\lb)}&1
\end{pmatrix}\begin{pmatrix}1&i\kappa_+(\lb)e^{\frac{2i}{\e}\phi_+(\lb)}\\
0&1
\end{pmatrix}.
\end{equation}
Because the first factor is analytic in a complex region
$\overline{\Omega_+}$ and the second in $\Omega_+$, this
factorization can be used to deform the jump contour: the interval
$(-1-\delta_1,u)$ can be deformed to a lens-shaped contour as shown
in Figure \ref{figure: sigmaS}.

\medskip

Define $S$ as follows,
\begin{equation}
S(\lambda)=\begin{cases}\begin{array}{ll}
T(\lambda)\begin{pmatrix}1&-i\kappa(\lb)e^{\frac{2i}{\e}\phi(\lb)}\\
0&1
\end{pmatrix},&\mbox{ in region I,}\\[3ex]
T(\lambda)\begin{pmatrix}1&0\\
i\kappa^*(\lb)e^{-\frac{2i}{\e}\phi(\lb)}&1
\end{pmatrix},&\mbox{ in region II},\\[3ex]
 T(\lambda), &\mbox{ elsewhere},
\end{array}
\end{cases}
\end{equation}
with $\kappa^*(\lb)=\bar{\kappa}(\bar{\lambda})$. Since $\kappa$
(resp. $\kappa^*$) is analytic in $\Omega_+$ (resp.
$\overline{\Omega_+}$), $S$ is analytic in each of the regions in
Figure \ref{figure: sigmaS} if we choose the lens sufficiently close
to the real line so that it lies in $\Omega$.

These are the RH conditions for $S$.
\subsubsection*{RH problem for $S$}
\begin{itemize}
\item[(a)] $S$ is analytic in $\mathbb C\setminus \Sigma_S$,
\item[(b)] $S_+(\lambda)=S_-(\lambda)v_S$ for $\lambda\in\Sigma_S$,
with
\begin{equation}\label{vS}
v_S(\lambda)=\begin{cases}
\begin{array}{lr}
\begin{pmatrix}1& i\kappa(\lb)e^{\frac{2i}{\e}\phi(\lb)}\\
0&1
\end{pmatrix},&\mbox{ on $\Sigma_1$},\\[3ex]
 \begin{pmatrix}1&0\\
i\kappa^*(\lb)e^{-\frac{2i}{\e}\phi(\lb)}&1
\end{pmatrix},&\mbox{ on $\Sigma_2$,}\\[3ex]
\begin{pmatrix}e^{-\frac{2i}{\e}\phi_+(\lb)}
&i\kappa(\lb)\\
i\bar\kappa(\lb)&(1-|r_0(\lb)|^2)e^{\frac{2i}{\e}\phi_+(\lb)}
\end{pmatrix},&\mbox{ as $\lambda\in (u,0)$,}\\[3ex]
v_T(\lambda),&\mbox{\hspace{-3cm} as
$\lambda\in(-\infty,-1-\delta_1)\cup(0,+\infty)$.}
\end{array}
\end{cases}
\end{equation}
\item[(c)] $S(\lambda)=\left(I+\bigO(\lambda^{-1})\right) \begin{pmatrix}1&1\\
i\sqrt{-\lambda}&-i\sqrt{-\lambda}\end{pmatrix}$ as
$\lambda\to\infty$.
\end{itemize}
For large $\lb$, $S(\lb)=T(\lb)$, and by (\ref{uT}) we have
\begin{equation}
u(x,t,\e)=u-2i\e\partial_x S_{11}^1(x,t,\e),\label{uS}
\end{equation}
where
\begin{equation}S_{11}(\lb;x,t,\e)=1+
\frac{S_{11}^1(x,t,\e)}{\sqrt{-\lb}}+O(\lb^{-1}),\qquad \mbox{ as
$\lb\to\infty$}.\end{equation} For $u=u(x,t_m)$ in the definition of
the $G$-function (\ref{def Godd}), we have $F(u;x,t_m)=0$ if $x$
belongs to the decreasing part of $u(x,t_m)$. For $t_m < t_m^c$ we
also have $F'(\xi;x,t_m)<0$ for $\xi\in(-1,0)$. If the contours
$\Sigma_1$ and $\Sigma_2$ are chosen sufficiently close to the real
line, one uses (\ref{phi0odd1}) to verify that, for any neighborhood
$\mathcal U$ of $u$, there is a $c>0$ such that
\begin{align}
\label{ineq phi1} &\Im\phi(\lambda;x,t_m;u)>c,&\mbox{for
$\lambda\in\Sigma_1\setminus \mathcal U$}, \\
\label{ineq phi2} &\Im\phi(\lambda;x,t_m;u)< -c,&\mbox{$\lambda\in\Sigma_2\setminus \mathcal U$},\\
\label{ineq phi3} &\Im\phi_+(\lambda;x,t_m;u)< -c,&\mbox{for
$\lambda\in(u,0)\setminus\mathcal U$}.
\end{align}
For $u=u^c=u(x^c,t_m^c)$, we have $F(u;x^c,t_m^c)=0$ and
$F'(\xi;x^c,t_m^c)<0$ for $\xi\in(-1,0)\setminus\{u\}$, and one
again uses (\ref{phi0odd}) to check that the inequalities (\ref{ineq
phi1})-(\ref{ineq phi3}) hold. Since $\phi$ is continuous in $x$ and
$t_m$, there must be a $\delta>0$ such that (\ref{ineq
phi1})-(\ref{ineq phi3}) hold also for $u=u^c$, and for
$|x-x^c|<\delta$, $|t_m-t_m^c|<\delta$ (possibly for a smaller
$c>0$). The inequalities imply together with
(\ref{r0}),(\ref{kappa1}), and (\ref{r2}) that the jump matrix $v_S$
for $S$ is uniformly close to constant matrices on
$\Sigma_S\setminus\mathcal U$ as $\e\to 0$: we have exponential
decay to $I$ for $\lambda<-1-\delta_1$ and on
$(\Sigma_1\cup\Sigma_2)\setminus\mathcal U$, and decay to
$i\sigma_1$ on $(u,0)\setminus\mathcal U$, with an error of order
$\bigO(\e)$.

\subsection{Outside parametrix}\label{section: outside}
We now deal with the jumps that do not tend to the identity matrix
as $\e\to 0$. Therefore we ignore for a moment all jumps that tend
to $I$ as $\e\to 0$, and a small neighborhood of $u$. We are then
left with a jump $i\sigma_1$ on $(u,0)$, and a jump $\sigma_1$ on
$(0,+\infty)$. We can explicitly construct a function with those
jump properties, and with the same asymptotic behavior as $S$.
Indeed, if we define
\begin{equation}
\label{def Pinfty}
P^{(\infty)}(\lambda)=(-\lambda)^{1/4}(u-\lambda)^{-\sigma_3
/4}\begin{pmatrix}1&1\\ i& -i
\end{pmatrix},
\end{equation}
we have
\subsubsection*{RH problem for $P^{(\infty)}$}
\begin{itemize}
\item[(a)] $P^{(\infty)}:\mathbb C\setminus [u, +\infty) \to
\mathbb C^{2\times 2}$ is analytic, \item[(b)] $P^{(\infty)}$
satisfies the following jump conditions on $(u, +\infty)$,
\begin{align}
&P_+^{(\infty)}=P_-^{(\infty)}\sigma_1, &\mbox{ as $\lambda\in (0, +\infty)$},\\
&P_+^{(\infty)}= iP_-^{(\infty)}\sigma_1, &\mbox{ as $\lambda\in
(u,0)$},\label{RHP Pinfty b}
\end{align}
\item[(c)]$P^{(\infty)}$ has the asymptotic behavior
\begin{equation}\label{RHP Pinfty c2}P^{(\infty)}(\lambda)=\left(I+\frac{u}{4\lambda}\sigma_3+\bigO(\lambda^{-2})\right)
\begin{pmatrix}1&1\\ i(-\lambda)^{1/2} & -i(-\lambda)^{1/2}\end{pmatrix},
\quad\mbox{as $\lb\to\infty$.}\end{equation}
\end{itemize}
This outside parametrix will determine the leading order asymptotic
behavior of $S$ as $\e\to 0$, but to prove this, we need to control
the jump matrices also in the vicinity of $u$. We will do this by
constructing a local parametrix near $u$.

\subsection{Local parametrix near $u$}\label{section: localu}

So far, there was no need to distinguish between the regular case,
where $x,t_m$ are fixed and where $u=u(x,t_m)$, and the singular
case, where $x,t_m$ are involved in a double scaling limit together
with $\e$ and where $u=u^c=u(x^c,t_m^c)$. The choices of $u$ will
however be crucial for the construction of the local parametrix. In
either case, we want to construct a local parametrix in a
neighborhood $\mathcal U$ of $u$ in such a way that
\subsubsection*{RH problem for $P$}
\begin{itemize}
\item[(a)]$P:\overline{\mathcal U}\setminus \Sigma_S\to\mathbb C^{2\times 2}$ is
analytic,
\item[(b)]$P$ satisfies the following jump condition on $\mathcal U \cap
\Sigma_S$,
\begin{equation}\label{RHP
P:b}P_+(\lambda)=P_-(\lambda)v_P(\lambda),
\end{equation}
with $v_P$ given by
\begin{equation}\label{vP}
v_{P}(\lb)=\begin{cases}
\begin{array}{ll}
\begin{pmatrix}1&ie^{\frac{2i}{\e}\phi(\lb;x,t_m)}\\
0&1
\end{pmatrix},&\mbox{ as $\lb\in\Sigma_1$},\\[3ex]
 \begin{pmatrix}1&0\\
ie^{-\frac{2i}{\e}\phi(\lb;x,t_m)}&1
\end{pmatrix},&\mbox{ as $\lb\in\Sigma_2$,}\\[3ex]
\begin{pmatrix}e^{-\frac{2i}{\e}\phi_+(\lb;x,t_m)}&i\\
i&0
\end{pmatrix},&\mbox{ as $\lambda\in (u^c,0)$,}
\end{array}
\end{cases}
\end{equation}
\item[(c)] As $\e\to 0$, we have the matching condition
\begin{equation}\label{matching}
P(\lambda)=(I+\bigO(\e^\gamma))P^{(\infty)}(\lambda),\qquad \mbox{
for $\lambda\in\partial \mathcal U$, \quad $\gamma>0$},
\end{equation}
between the local parametrix and the outside parametrix.
\end{itemize}
Such a parametrix has, for $\e\to 0$, approximately the same jumps
as $S$ has on $\mathcal U\cap \Sigma_S$. The idea is that $P$ will
approximate $S$ near $u$ for small $\e$, whereas $P^{(\infty)}$ will
be a good approximation elsewhere.

Note that by (\ref{phi0odd}), $\phi(u;x,t_m)$ behaves like
$c(u-\lambda)^{3/2}$ as $\lambda\to u$ for $t_m<t_m^c$, but
$\phi(u^c,x^c,t_m^c)$ behaves like $c(u^c-\lambda)^{7/2}$ as
$\lambda\to u^c$, so the  jump matrix $v_P$ behaves differently in
the regular case than in the singular case in the vicinity of $u$.
This is the reason why we need to build different local parametrices
in both cases. In the regular case, we will be able to construct a
local parametrix with the value of $\gamma$ in (\ref{matching})
equal to $1$, in the singular case we will have a weaker matching
with $\gamma=1/7$.
\begin{figure}[t]
\begin{center}
    \setlength{\unitlength}{1mm}
    \begin{picture}(137.5,26)(22,11.5)

        \put(85,25){\thicklines\circle*{.8}} \put(80,25){$u$}
        \put(85,25){\line(1,0){45}}
        \put(124,25){\thicklines\vector(1,0){.0001}}
        \qbezier(85,25)(65,45)(45,50) \put(65,41.1){\thicklines\vector(4,-3){.0001}}
        \qbezier(85,25)(65,5)(45,0) \put(65,8.8){\thicklines\vector(4,3){.0001}}

    \end{picture}\vspace{1cm}
    \caption{The jump contour for $P$}
    \label{Figure: contour P}
\end{center}
\end{figure}
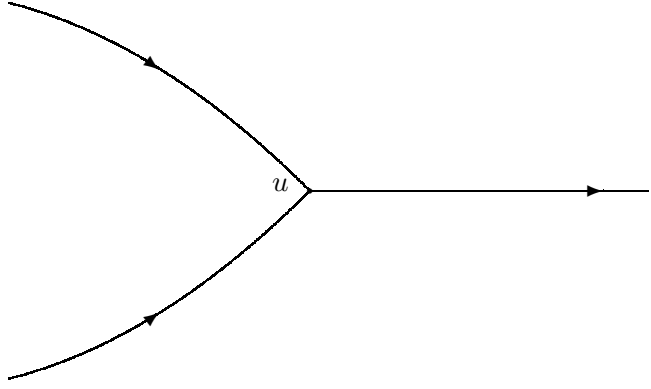

\subsubsection{The regular case: $t_m<t_m^c$}

Here we let $u=u(x,t_m)$ be defined by the equation (\ref{char}),
i.e.\ $u(x,t_m)$ is the solution to the unperturbed equation
(\ref{mBurgers}). Then we will construct the local parametrix near
$u$ using the following model RH problem.

\subsubsection*{RH problem for $\Phi$}
\begin{itemize}
    \item[(a)] $\Phi$ is analytic for $\zeta\in\mathbb{C}
    \setminus \Gamma$, with $\Gamma=\{\zeta\in\mathbb C: \arg\zeta=0, \arg\zeta=\frac{2\pi}{3},
    {\rm or}  \arg\zeta=-\frac{2\pi}{3},\}$.
    \item[(b)] $\Phi$ satisfies the jump relations
    \begin{align}
        \label{RHP Phi: b1}
        &\Phi_+(\zeta)=\Phi_-(\zeta)\begin{pmatrix}
            e^{-2\theta(\zeta)} & i \\
            i & 0
        \end{pmatrix}
        ,& \mbox{for $\arg\zeta=0$,} \\[1ex]
        \label{RHP Phi: b2}
        &\Phi_+(\zeta)=\Phi_-(\zeta)\begin{pmatrix}
            1 & ie^{2\theta(\zeta)} \\
            0 & 1
        \end{pmatrix}
        ,& \mbox{for $\arg\zeta=\frac{2\pi}{3}$.}
        \\[1ex]
        \label{RHP Phi: b3}
        &\Phi_+(\zeta)=\Phi_-(\zeta)
        \begin{pmatrix}
            1 & 0 \\
            ie^{2\theta(\zeta)} & 1
        \end{pmatrix},& \mbox{for $\arg\zeta=-\frac{2\pi}{3}$,}
    \end{align}
    where $\theta(\zeta)=\frac{2}{3}\zeta^{3/2}$, with branch cut on $(-\infty,0)$.
    \item[(c)] $\Phi$ has the following behavior as $\zeta\to\infty$,
    \begin{equation}\label{RHP Phi: c}
        \Phi(\zeta)=\frac{1}{\sqrt
        2}(-\zeta)^{-\frac{1}{4}\sigma_3}\begin{pmatrix}1&1\\-1&1\end{pmatrix}\left(I        +\bigO(\zeta^{-3/2})\right),
    \end{equation}
    with the branch cuts of $(-\zeta)^{\pm\frac{1}{4}}$ along $(0,+\infty)$.
\end{itemize}
Several RH problems equivalent to this one have appeared in the
literature, see e.g.\ \cite{DKMVZ2, DKMVZ1, Deift, DVZ}, and it is
well-known that this problem can be solved explicitly in terms of
the Airy function. The unique solution to this precise problem has
been constructed in \cite{CG2}. In the singular case later on, we
will use the same model RH problem, but with a different value for
$\theta$.  We only need the existence of a $2\times 2$ matrix-valued
function $\Phi$ satisfying the above RH conditions, the precise
construction in terms of the Airy function is irrelevant.

\medskip

We define the parametrix $P=P(\lambda;x,t_m,\e)$ by
\begin{equation}\label{definition hatP}
P(\lambda)=E(\lambda;\epsilon)\Phi(\epsilon^{-2/3}f(\lambda)),
\end{equation}
where $E$ is an analytic function in $\mathcal U$ and $f$ is a
conformal map from $\mathcal U$ to a neighborhood of $0$. Let us
first define $f$ by the requirement that
\begin{equation}\label{def f}
\theta(f(\lambda))=\frac{2}{3}f(\lambda)^{3/2}=\widehat\phi(\lambda),
\end{equation}
and \begin{equation}
\widehat\phi(\lambda)=\pm i\phi(\lambda), \qquad \mbox{ for $\pm\Im\lambda>0$,}
\end{equation}
so that the branch cut of $\widehat\phi$ coincides with the one of
$f^{3/2}$ along $\lambda<u$. By (\ref{phi0odd}), equation (\ref{def
f}) only defines $f$ analytically if $F(u;x,t_m)=0$, which is true
because of our choice of $u$. Then we have
\begin{equation}
f(u)=0, \qquad f'(u)=F'(u;x,t_m)^{2/3}>0.
\end{equation}
Now we can choose the lens $\Sigma_S$ in $\mathcal U$ in such a way
that $f$ maps $\Sigma_S\cap\mathcal U$ to the jump contour $\Gamma$
for $\Phi$. Then $P$ has jumps on $\Sigma_S\cap \mathcal U$ which
are given by (\ref{vP}).

\medskip

Now define $E$ by
\begin{equation}
E(\lambda;\epsilon)=\frac{1}{\sqrt
2}P^{(\infty)}(\lambda)\begin{pmatrix}1&-1\\1&1\end{pmatrix}
(-\epsilon^{-2/3}f(\lambda))^{\frac{\sigma_3}{4}}.
\end{equation}
Then it is easily verified that $E$ is analytic in $\mathcal U$, and using (\ref{RHP Phi: c}), we obtain the matching
\begin{equation}\label{matchingregular}
P(\lambda)P^{(\infty)}(\lambda)^{-1}=P^{(\infty)}(\lambda)\left(I+\bigO(\e)\right)
P^{(\infty)}(\lambda)^{-1}=I+\bigO(\e),
\end{equation}
for $\lambda\in\partial U$ as $\e\to 0$. When $x,t_m$ approach $x^c,
t_m^c$, the uniform convergence breaks down because $f'(u)$
approaches zero. Therefore we will now construct a different local
parametrix in the singular case.

\subsubsection{The singular case: double scaling limit $\e\to 0$, $x\to x^c$, $t_m\to t_m^c$}
In this case we need a slightly modified model RH problem, which is
obtained by replacing $\theta$ in the RH problem for $\Phi$ by
\begin{equation}
\theta^c(\zeta;X,T)=\frac{1}{105}\zeta^{7/2}-\frac{T}{3}\zeta^{3/2}+X\zeta^{1/2}.
\end{equation}
We then obtain the RH problem

\subsubsection*{RH problem for $\Phi^c$}
\begin{itemize}
    \item[(a)] $\Phi^c$ is analytic for $\zeta\in\mathbb{C}
    \setminus \Gamma^c$, with $\Gamma^c=\{\zeta\in\mathbb C: \arg\zeta=0, \arg\zeta=\frac{6\pi}{7}, {\rm or}  \arg\zeta=-\frac{6\pi}{7},\}$.
    \item[(b)] $\Phi^c$ satisfies the jump relations
    \begin{align}
        \label{RHP Phi: b1c}
        &\Phi_+^c(\zeta)=\Phi_-^c(\zeta)\begin{pmatrix}
            e^{-2\theta^c(\zeta)} & i \\
            i & 0
        \end{pmatrix}
        ,& \mbox{for $\arg\zeta=0$,} \\[1ex]
        \label{RHP Phi: b2c}
        &\Phi_+^c(\zeta)=\Phi_-^c(\zeta)\begin{pmatrix}
            1 & ie^{2\theta^c(\zeta)} \\
            0 & 1
        \end{pmatrix}
        ,& \mbox{for $\arg\zeta=\frac{6\pi}{7}$.}
        \\[1ex]
        \label{RHP Phi: b3c}
        &\Phi_+^c(\zeta)=\Phi_-^c(\zeta)
        \begin{pmatrix}
            1 & 0 \\
            ie^{2\theta^c(\zeta)} & 1
        \end{pmatrix},& \mbox{for $\arg\zeta=-\frac{6\pi}{7}$.}
    \end{align}
    \item[(c)] $\Phi^c$ has the following behavior as $\zeta\to\infty$,
    \begin{multline}\label{RHP Phi: cc}
        \Phi^c(\zeta)=\frac{1}{\sqrt
        2}(-\zeta)^{-\frac{1}{4}\sigma_3}\begin{pmatrix}1&1\\-1&1\end{pmatrix}\left(I+iQ\sigma_3(-\zeta)^{-1/2}-\frac{1}{2}
        \begin{pmatrix}Q^2&U\\U&Q^2\end{pmatrix}(-\zeta)^{-1}\right. \\
        \left.
        -\frac{i}{2}(-\zeta)^{-3/2}\begin{pmatrix}W&V\\
        -V&-W\end{pmatrix}+\bigO(\zeta^{-2})\right).
    \end{multline}
\end{itemize}

In \cite{CV1}, existence of a solution for a slightly different but
equivalent RH problem has been proved, and the transformation to the
RH problem for $\Phi^c$ has been given in \cite{CG}. We write
$\Phi^c=\Phi^c(\zeta;X,T)$ for the solution to this RH problem. In
the regular case, the constants $Q, U$ vanished, but this is no
longer true in the singular case: they now depend in a
transcendental way on the parameters $X,T$. We have $Q=Q(X,T)$,
$U=U(X,T)$,  where $U$ solves the ${\rm P_I^2}$ equation with
asymptotics given by (\ref{PI2asym}). Furthermore we have the
relations
\begin{align}
&\label{Q}Q_X=U,\\
&\label{S}W_X=Q^2U+U^2,\\
&\label{V}V=-QU-\frac{1}{2}U_X.
\end{align}
Those identities follow from the fact that $\Phi_X^c\Phi^{c,-1}$ is a
polynomial in $\zeta$. Substituting (\ref{RHP Phi: cc}) gives a
function with terms proportional to $\zeta^{-1}, \zeta^{-2}, \ldots$
as $\zeta\to\infty$. The requirement that those terms vanish, leads
to (\ref{Q})-(\ref{V}). Using the equation (\ref{PI20}) and the
asymptotics for $Q$ obtained in \cite{CV1}, it is straighforward to
derive the identity
\begin{equation}
\label{QID}
Q=\dfrac{1}{240}U_XU_{XXX}-\dfrac{U_{XX}^2}{480}+XU-\dfrac{T}{2}U^2+\dfrac{U^4}{24}+\dfrac{1}{24}UU_X^2.
\end{equation}
For any $X_0,T_0\in\mathbb R$, there are complex neighborhoods of
$X_0$ and $T_0$ such that the RH problem is solvable for $X,T$ in
those neighborhoods, and such that (\ref{RHP Phi: c}) holds
uniformly in $X,T$.

We want to construct a parametrix of the form
\begin{equation}\label{definition Pc}
P(\lambda)=E(\lambda;\epsilon)\Phi^c\left(\epsilon^{-2/7}f(\lambda);
\e^{-6/7}g_1(\lambda;x,t_m),\e^{-4/7}g_2(\lambda;t_m)\right).
\end{equation}
where $f, g_1, g_2, E$ are analytic functions which are to be
determined. If we want $P$ to satisfy the jump conditions given in
(\ref{RHP P:b}), we need to construct $f, g_1, g_2$ in such a way
that
\begin{equation}\label{def f g1}
\theta^c\left(\e^{-2/7}f(\lambda);\e^{-6/7}g_1(\lambda;x,t_m),\e^{-4/7}g_2(\lambda;t_m)\right)=\frac{1}{\e}\widehat\phi(\lambda),
\end{equation}
or equivalently
\begin{equation}\label{def f g}
\theta^c\left(f(\lambda);g_1(\lambda;x,t_m),g_2(\lambda;t_m)\right)=\widehat\phi(\lambda).
\end{equation}
Moreover we want that
$\e^{-6/7}g_1(\lambda;x,t_m),\e^{-4/7}g_2(\lambda;t_m)$ remain
bounded in the double scaling limit where $\e\to 0$, $x\to x^c$,
$t_m\to t_m^c$ for fixed $\e>0$. This is the reason why we need to
restrict to the scalings (\ref{double scaling}).

\medskip

If we take $u=u^c=u(x^c,t_m^c)$, we have
\[F(u^c;x^c,t_m^c)=F'(u^c;x^c,t_m^c)=F''(u^c;x^c,t_m^c)=0,\] and
(\ref{phi0odd}) implies
\begin{multline}\label{phic}
\widehat\phi(\lambda;x,t_m,\e)=-\sqrt{\lambda-u^c}
(F(u^c;x,t_m)-F(u^c;x^c,t_m^c))\\-\dfrac{2}{3}(\lambda-u^c)^{\frac{3}{2}}(F'(u^c;x,t_m)-F'(u^c;x^c,t_m^c))
-\dfrac{4}{15}(\lambda-u^c)^{\frac{5}{2}}(F''(u^c;x,t_m)-F''(u^c;x^c,t_m^c))
\\-\dfrac{4}{15}\int_{u^c}^{\lambda}F'''(\xi;x,t_m)(\lambda-\xi)^{\frac{5}{2}}d\xi.
\end{multline}
Then it is clear that the first line in (\ref{phic}) vanishes like a
square root as $\lambda\to u^c$, the second line vanishes like
$c(\lambda-u^c)^{3/2}$, and the third line behaves like
$c'(\lambda-u^c)^{7/2}$. Therefore we define $f, g_1, g_2$ by the
equations
\begin{align}
&\frac{1}{105}f(\lambda)^{7/2}=-\dfrac{4}{15}\int_{u^c}^{\lambda}F'''(\xi;x^c,t_m^c)(\lambda-\xi)^{\frac{5}{2}}d\xi\\[1ex]
&g_1(\lambda;x,t_m)f(\lambda)^{1/2}=-\sqrt{\lambda-u^c}
(F(u^c;x,t_m)-F(u^c;x^c,t_m^c)),\\[1ex]
&-\frac{g_2(\lambda;t_m)}{3}f(\lambda)^{3/2}=-\dfrac{2}{3}(\lambda-u^c)^{\frac{3}{2}}(F'(u^c;x,t_m)-F'(u^c;x^c,t_m^c))\nonumber\\
&\qquad\qquad
-\dfrac{4}{15}(\lambda-u^c)^{\frac{5}{2}}(F''(u^c;x,t_m)-F''(u^c;x^c,t_m^c))\nonumber\\
&\qquad\qquad\qquad\qquad-\dfrac{4}{15}\int_{u^c}^{\lambda}(F'''(\xi;x,t_m)-F'''(\xi;x^c,t_m^c))
(\lambda-\xi)^{\frac{5}{2}}d\xi.
\end{align}
One verifies that this defines $f, g_1, g_2$ analytically in
$\mathcal U$ and moreover in such a way that (\ref{def f g}) holds.
A straightforward calculation yields
\begin{align}
&\label{f1}f(u^c)=0, \qquad f'(u^c)=(8k)^{\frac{2}{7}},\qquad
f''(u^c)=-\dfrac{64}{63}\dfrac{F^{(4)}(u^c)}{(8k)^{5/7}},
\\
&\label{g11}g_1(u^c;x,t_m)=\dfrac{\tilde{x}-\tilde{x}^c}{(8k)^{1/7}},\qquad
g_1'(u^c,x,t_m)=-\dfrac{f''(u^c)}{4(8k)^{2/7}
}g_1(u^c,x,t_m),\\
&\label{g21}g_2(u^c;t_m)=\dfrac{2mC_m(u^c)^{m-1}(t_m-t^c_m)}{(8k)^{3/7}},\\
&\label{g22}g_2'(u^c;t_m)=\left(\frac{2(m-1)}{5u^c}-\frac{3f''(u^c)}{4f'(u^c)}\right)g_2(u^c;t_m),
\end{align}
where \begin{equation} \label{tildex} \tilde{x}=x-C_m(u^c)^mt_m,
\qquad k=-F'''(u^c)>0.
\end{equation}
Similarly as in the regular case, we can choose the lens in such a
way that $f(\Sigma_S\cap\mathcal U)\subset \Gamma$. Then $P$
satisfies the jump condition (\ref{RHP P:b}) for
$\lambda\in\Sigma_S\cap \mathcal U$.

\medskip

In order to have a good matching between $P$ and $P^{(\infty)}$ on
$\partial\mathcal U$, we define the analytic pre-factor $E$ by
\begin{equation}
E(\lambda;\epsilon)=\frac{1}{\sqrt
2}P^{(\infty)}(\lambda)\begin{pmatrix}1&-1\\1&1\end{pmatrix}
(-\epsilon^{-2/7}f(\lambda))^{\frac{\sigma_3}{4}}.
\end{equation}
Using the definition (\ref{def Pinfty}) of $P^{(\infty)}$, one
checks directly that $E$ is analytic in $\mathcal U$. For
$\lambda\in\partial\mathcal U$ and $\e^{-4/7}g_1(\lb;x,t_m)$,
$\e^{-6/7}g_2(\lb;t_m)$ in small complex neighborhoods of
$X,T\in\mathbb R$, we have the matching
\begin{multline}\label{RHP P:c2}
P(\lambda)P^{(\infty)}(\lambda)^{-1}=P^{\infty}(\lambda)\left(I+iQ\sigma_3(-f(\lambda))^{-1/2}\epsilon^{1/7}\right.\\
\left.-\frac{1}{2}
        \begin{pmatrix}Q^2&U\\U&Q^2\end{pmatrix}(-f(\lambda))^{-1}\epsilon^{2/7}-\frac{i}{2}
        \begin{pmatrix}W&V\\-V&-W\end{pmatrix}(-f(\lambda))^{-3/2}\epsilon^{3/7}
                +\bigO(\epsilon^{4/7})\right)
        P^{\infty}(\lambda)^{-1},
\end{multline}
as $\e\to 0$, with
\begin{align}
&\label{abbU}U=U(\epsilon^{-6/7}g_1(\lb;x,t_m),\epsilon^{-4/7}g_2(\lb;t_m)),
\\
&\label{abbQ}Q=Q(\epsilon^{-6/7}g_1(\lb;x,t_m),\epsilon^{-4/7}g_2(\lb;t_m)),\\
&\label{abbW}W=W(\epsilon^{-6/7}g_1(\lb;x,t_m),\epsilon^{-4/7}g_2(\lb;t_m)),\\
&\label{abbV}V=V(\epsilon^{-6/7}g_1(\lb;x,t_m),\epsilon^{-4/7}g_2(\lb;t_m)).
\end{align}
In the double scaling limit given by (\ref{double scaling}), we have
by (\ref{g11})-(\ref{g21}) that \[\e^{-4/7}g_1(u^c;x,t_m)\to
T,\qquad \e^{-6/7}g_2(u^c;t_m)\to X,\] and from the definitions of
$g_1$ and $g_2$, it follows that
$\e^{-4/7}g_1(\lb;x,t_m),\e^{-6/7}g_2(\lb;t_m)$ lie in a small
neighborhood of $X,T$ if $\lambda$ is sufficiently close to $u^c$.
Summarizing, the matching (\ref{RHP P:c2}) holds in the double
scaling limit if we have chosen $\mathcal U$ small enough.

\subsection{Final transformation $S\mapsto R$}

We define
\begin{align}\label{def R}&
R(\lambda)=\begin{cases}S(\lambda)P^{(\infty)}(\lambda)^{-1},
& \mbox{ as $\lambda\in \mathbb C\setminus \mathcal U$,}\\
S(\lambda)P(\lambda)^{-1}, & \mbox{ as $\lambda\in \mathcal U$.}
\end{cases}
\end{align}
The outside parametrix has been constructed in such a way that the
jumps of $R$ are close to $I$ as $\e\to 0$. On $(0,+\infty)$, the
jump for $S$ cancels out exactly with the jump of $P^{(\infty)}$. On
the other parts of $\Sigma_S\setminus\overline{\mathcal U}$, we have
\begin{eqnarray}
R_-^{-1}(\lambda)R_+(\lambda)&=&
P_-^{(\infty)}(\lambda)v_S(\lambda)v_{P^{(\infty)}}^{-1}(\lambda)P_-^{(\infty)}(\lambda)^{-1}\\
&=&I+\bigO(\e),\qquad\qquad\qquad\qquad\mbox{ as $\e\to
0$}.\end{eqnarray} For $\lambda\in\mathcal U\cap\Sigma_S$, we have
\begin{equation}R_-^{-1}(\lambda)R_+(\lambda)=P_-(\lambda)v_S(\lambda)v_P^{-1}(\lambda)P_-^{-1}(\lambda).\end{equation}
On one hand it follows from the construction of the parametrix that
$P_-(\lambda)$ is uniformly bounded for $\lambda\in\mathcal
U\cap\Sigma_S$. On the other hand
\begin{equation}
v_Sv_P^{-1}=\begin{cases}
\begin{array}{ll}
\begin{pmatrix}1&i(\kappa -1)e^{\frac{2i}{\e}\phi}\\
0&1
\end{pmatrix},&\mbox{ on $\Sigma_1\cap \mathcal U$},\\[3ex]
 \begin{pmatrix}1&0\\
i(\kappa^* -1)e^{-\frac{2i}{\e}\phi}&1
\end{pmatrix},&\mbox{ on $\Sigma_2\cap \mathcal U$,}\\[3ex]
\begin{pmatrix}\kappa&i(\kappa-1)e^{-\frac{2i}{\e}\phi_+}\\
-i(1-|r|^2)e^{\frac{2i}{\e}\phi_+}&\kappa^*+(1-|r|^2)
\end{pmatrix},&\mbox{ on $(u^c,0)\cap \mathcal U$.}
\end{array}
\end{cases}
\end{equation}
Except for the $21$-entry on $(u^c,0)$, the exponentials in the
above matrices are uniformly bounded on the jump contours inside
$\mathcal U$. The asymptotics (\ref{kappa1}) and (\ref{r2}) for
$\kappa$ and $r$ ensure that $v_Sv_P^{-1}=I+\bigO(\e)$. We obtain
the following RH problem for $R$.

\subsubsection*{RH problem for $R$}
\begin{itemize}
\item[(a)]$R$ is analytic in $\mathbb C\setminus\Sigma_R$, with
$\Sigma_R=(\Sigma_S\cup\partial\mathcal U)\setminus (0,+\infty)$ as shown in Figure \ref{figure: sigmaR},
\item[(b)]$R_+(\lambda)=R_-(\lambda)v_R(\lambda)$, where the jump matrix $v_R$
is given by
\begin{align}\label{vR}
&
v_R(\lambda)=\begin{cases}P(\lambda)P^{(\infty)}(\lambda)^{-1},&\mbox{
as $\lambda\in \mathbb \partial \mathcal U$,}\\
I+\bigO(\epsilon),&\mbox{ as $\lambda\in \Sigma_R\setminus \partial
\mathcal U$ as $\e\to 0$.}
\end{cases}
\end{align}
\item[(c)]$R(\lambda)=I+\bigO(\lambda^{-1})$ as $\lambda\to\infty$.
\end{itemize}
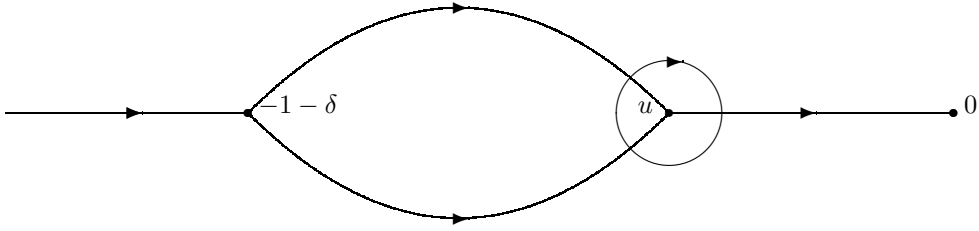
\begin{figure}[t]
\begin{center}

    \setlength{\unitlength}{1.4mm}
    \begin{picture}(137.5,26)(22,11.5)
        \put(112,25){\thicklines\circle*{.8}}
        \put(45,25){\thicklines\circle*{.8}}
        \put(22,25){\line(1,0){23}}
        \put(35,25){\thicklines\vector(1,0){.0001}}
        \put(85,25){\circle{15}}
\put(86.3,29.9){\thicklines\vector(1,0){.0001}}
        \put(112,25){\thicklines\circle*{.8}}
        \put(113,25){\small $0$}
        \put(45,25){\thicklines\circle*{.8}}
        \put(46,25){\small $-1-\delta$}
        \put(85,25){\thicklines\circle*{.8}} \put(82,25){\small $u$}
        \put(99,25){\thicklines\vector(1,0){.0001}}
        \put(85,25){\line(1,0){27}}
        \put(22,25){\line(1,0){23}}
        \put(35,25){\thicklines\vector(1,0){.0001}}
        \qbezier(45,25)(65,45)(85,25) \put(66,35){\thicklines\vector(1,0){.0001}}
        \qbezier(45,25)(65,5)(85,25) \put(66,15){\thicklines\vector(1,0){.0001}}
    \end{picture}
    \caption{The contour $\Sigma_R$ after the final
        transformation $S\mapsto R$.}
    \label{figure: sigmaR}
\end{center}
\end{figure}

From the matching (\ref{matchingregular}) in the regular case where
$t_m<t_m^c$, it follows that
\begin{equation}
v_R(\lambda)=I+\bigO(\e),\qquad \mbox{ as $\e\to 0$,}
\end{equation}
uniformly for $\lambda\in\Sigma_R$, since $P^{(\infty)}$ is bounded
on $\partial\mathcal U$. In the singular case, we use (\ref{RHP
P:c2}) to conclude that
\begin{equation}
v_R(\lambda)=I+\bigO(\e^{1/7}),
\end{equation}
uniformly for $\lambda\in\Sigma_S$, in the double scaling limit
where $\e\to 0$, $x\to x^c$, and $t_m\to t_m^c$ in such a way that
(\ref{double scaling}) holds.

\medskip

Following the general theory for small-norm RH problems
\cite{DKMVZ1}, one shows that
\begin{equation}\label{intR}
R(\lambda)=I+\frac{1}{2\pi i}\int_{\Sigma_R}R_-(s)(v_R(s)-I)\frac{ds}{s-\lambda},\qquad \mbox{ for $\lambda\in\mathbb C\setminus R$,}
\end{equation}
and that
\begin{equation}
\|R_- - I\|_{L^2(\Sigma_R)}=\bigO(\e^\gamma),
\end{equation}
where $\gamma=1$ in the regular case as $\e\to 0$, and $\gamma=1/7$ in the double scaling limit in the singular case.

\section{Proof of the results}\label{section: proof}

\subsection{Proof of Theorem \ref{theorem: 1}}\label{section:
theorem1}

Note that (\ref{intR}) implies that
\begin{equation}\label{asRlambda}
R(\lambda)=I+\frac{R_1}{\lambda}+\bigO(\lambda^{-2}),\qquad\mbox{ as
$\lambda\to\infty$,}
\end{equation}
with
\begin{equation}
R_1=R_1(x,t_m,\e)=-\frac{1}{2\pi i}\int_{\Sigma_R}R_-(s)(v(s)-I)ds.
\end{equation}
This gives
\begin{equation}\label{formulaR1}
R_1=-\frac{1}{2\pi
i}\int_{\Sigma_R}(R_-(s)-I)(v(s)-I)ds-\frac{1}{2\pi
i}\int_{\Sigma_R}(v_R(s)-I)ds=\bigO(\e),
\end{equation}
as $\e\to 0$ by the Cauchy-Schwarz inequality. Differentiating the
integral equation (\ref{intR}) and (\ref{formulaR1}) in $x$, one
similarly obtains the estimates
\begin{equation}
\|R_{x,-}\|_{L^2(\Sigma_R)}=\bigO(\e),\qquad \partial_x
R_{1}=\bigO(\e),
\end{equation}
using the fact that $\|v_{R,x}\|_{L^2(\Sigma_R)}=\bigO(\e)$.
  Now by
(\ref{def R}), we have that
$S(\lambda)=R(\lambda)P^{(\infty)}(\lambda)$ for large $\lambda$,
which implies by (\ref{RHP Pinfty c2}) that
\begin{equation}
S_{11}(\lambda)=1-i\frac{R_{1,12}}{(-\lambda)^{1/2}}+\bigO(\lambda^{-1}),
\qquad\mbox{ as $\lambda\to\infty$}.
\end{equation}
By (\ref{uS}), we obtain
\begin{equation}
\label{uRa} u(x,t_m,\epsilon)=u(x,t_m)-2\epsilon\partial_x
R_{1,12}(\tilde{x},t_m,\epsilon)=u(x,t_m)+\bigO(\e^2),\end{equation}
which proves Theorem \ref{theorem: 1} in the case where
$x<x_M+C_mt_m$.

\begin{remark}\label{remark: increasing}
If $x=x_M+C_mt_m$, the RH analysis remains the same, except for the
fact that $u=-1$ and $F'(u;x,t_m)=+\infty$. This implies that the
jump matrix for $S$ decays to $I$ also near $u$, and there is no
need to construct a local parametrix. If $x>x_M+C_mt_m$, we have
that $-x+C_mu^{m-1}t_m+f_R(u)=0$, with $f_R$ the inverse of the
increasing part of $u_0$, but $F(u;x,t_m)=-x+C_mu^{m-1}t_m+f_L(u)$
does not vanish (in general). Because of this, we need to modify the
$G$-function:
\[
G_m(\lambda;x,t_m;u)=\frac{\sqrt{u - \lambda}}{\pi}\left[
\int_{u}^0\frac{F_A(\eta;x,t_m)}{(\eta
-\lambda)\sqrt{\eta-u}}d\eta-\int_{-1}^u\frac{\tau(\eta)}{(\eta
-\lambda)\sqrt{u-\eta}}d\eta\right],
\]
with
 \begin{equation*}
 \tau(\lambda)=\int_{f_L(\lambda)}^{f_{R}(\lambda)}\sqrt{\lambda-u_0(x)}dx,\qquad
 \mbox{ for $-1<\lambda<0$.}
  \end{equation*}
This gives the additional jump condition
\begin{equation*}
G_{m,+}(\lambda)- G_{m,-}(\lb)=-2i\tau(\lambda),\qquad\mbox{for
$\lambda\in (-1,u)$}.
\end{equation*}
Then $\phi$ takes the form
\begin{multline*}
\phi(\lambda;x,t_m)=-i\tau(\lambda)-\sqrt{u-\lambda}
F(u;x,t_m)+\dfrac{2}{3}(u-\lambda)^{\frac{3}{2}}F'(u;x,t_m)\\
-\dfrac{4}{15}(u-\lambda)^{\frac{5}{2}}F''(u;x,t_m)
-\dfrac{4}{15}\int_{u}^{\lambda}F'''(\xi;x,t_m)(\xi-\lambda)^{\frac{5}{2}}d\xi.
\end{multline*}
With those modifications, the RH analysis can be carried on
similarly as before, see also \cite{DVZ}.
\end{remark}

\subsection{Proof of Theorem \ref{theorem: 2}}
Using (\ref{RHP P:c2}), we can expand the jump matrix $v_R$ in
fractional powers of $\e$ in the double scaling limit,
\begin{equation}\label{vRexpansion}
v_R(\lambda)=I+\epsilon^{1/7}\Delta^{(1)}(\lambda)+\epsilon^{2/7}\Delta^{(2)}(\lambda)+\epsilon^{3/7}
\Delta^{(3)}(\lambda)+ \bigO(\epsilon^{4/7}),
\end{equation}
with
\begin{align}\label{delta1}
&\Delta^{(1)}(\lambda)=iQ \cdot
(-f(\lambda))^{-1/2}P^{(\infty)}(\lambda)\sigma_3
P^{(\infty)}(\lambda)^{-1},\\
\label{delta2}
&\Delta^{(2)}(\lambda)=-\frac{1}{2}(-f(\lambda))^{-1}P^{(\infty)}(\lambda)
\begin{pmatrix}Q^2&U\\U&Q^2\end{pmatrix}
P^{(\infty)}(\lambda)^{-1},\\
\label{delta3}
&\Delta^{(3)}(\lambda)=-\frac{i}{2}
(-f(\lambda))^{-3/2}P^{(\infty)}(\lambda)\begin{pmatrix}W&V\\-V&-W\end{pmatrix}
P^{(\infty)}(\lambda)^{-1},
\end{align}
for $\lambda\in\partial \mathcal U$, and
\begin{equation}
\Delta^{(1)}(\lambda)=\Delta^{(2)}(\lambda)=\Delta^{(3)}(\lambda)=0,
\qquad \mbox{for $\lambda\in\Sigma_R\setminus\partial\mathcal U$,}
\end{equation}
 since the jump matrices are equal to $I$ up to an error of $\bigO(\e)$ on the other parts of the contour.
Note that the functions $\Delta^{(1)}$ and $\Delta^{(2)}$ are
meromorphic functions in $\mathcal U$ with simple poles at $u^c$,
and that $\Delta^{(3)}$ is meromorphic in $\mathcal U$ with a double
pole at $u^c$. Observe that the $x$-derivative of $v_R$ is not close
to $I$ on $\partial \mathcal U$ as $\e\to 0$, since the
$x$-derivatives of $U$ and $Q$ cause multiplication with $\e^{-6/7}$
by (\ref{abbU})-(\ref{abbQ}).

\medskip

Substituting (\ref{vRexpansion}) in (\ref{intR}) yields an
asymptotic expansion for the RH solution $R$ of the form
\begin{equation}\label{Rexpansionepsilon}
R(\lambda)=I+\epsilon^{1/7}R^{(1)}(\lambda)+\epsilon^{2/7}R^{(2)}(\lambda)+\epsilon^{3/7}R^{(3)}(\lambda)
+\bigO(\epsilon^{4/7}).
\end{equation}
Combining (\ref{vRexpansion}) with (\ref{Rexpansionepsilon}) and the
jump relation $R_+(\lambda)=R_-(\lambda)v_R(\lambda)$ gives the
following relations for $\lambda\in\partial \mathcal U$,
\begin{align}\label{RHPR1}
& R_+^{(1)}(\lambda)=R_-^{(1)}(\lambda) + \Delta^{(1)}(\lambda),\\
\label{RHPR2}& R_+^{(2)}(\lambda)=R_-^{(2)}(\lambda) +
R_-^{(1)}(\lambda)\Delta^{(1)}(\lambda)+ \Delta^{(2)}(\lambda)\\
\label{RHPR3}& R_+^{(3)}(\lambda)=R_-^{(3)}(\lambda) +
R_-^{(1)}(\lambda)\Delta^{(2)}(\lambda)+
R_-^{(2)}(\lambda)\Delta^{(1)}(\lambda)+\Delta^{(3)}(\lambda).
\end{align}
In addition we know that $R(\lambda)\to I$ as $\lambda\to\infty$,
and consequently $R^{(j)}(\lambda)\to 0$ for $j=1,2,3$. So we have
additive jump relations and asymptotic conditions for $R^{(1)}$,
$R^{(2)}$, and $R^{(3)}$, and it is easily verified that those
conditions determine $R^{(1)}$, $R^{(2)}$, and $R^{(3)}$ uniquely.
We obtain
\begin{align}\label{R1}
& R^{(1)}(\lambda)= \begin{cases}\frac{1}{\lambda
-u^c}\Res(\Delta^{(1)};u^c),
&\mbox{ as $\lambda\in\mathbb C\setminus \overline{\mathcal U}$}\\
\frac{1}{\lambda -u^c}\Res(\Delta^{(1)};u^c)-\Delta^{(1)}(\lambda),
&\mbox{ as $\lambda\in \mathcal U$,}\end{cases}\\
\label{R2} & R^{(2)}(\lambda)= \nonumber\\&\begin{cases} \frac{1}{\lambda
-u^c}\Res(R^{(1)}\Delta^{(1)}+\Delta^{(2)};u^c),
&\mbox{ as $\lambda\in\mathbb C\setminus \overline{\mathcal U}$,}\\
\frac{1}{\lambda
-u^c}\Res(R^{(1)}\Delta^{(1)}+\Delta^{(2)};u^c)-R^{(1)}\Delta^{(1)}(\lambda)
-\Delta^{(2)}(\lambda),
&\mbox{ as $\lambda\in \mathcal U$.}\\
\end{cases}
\end{align}
After a straightforward calculation we find using (\ref{delta1}),
(\ref{delta2}), and (\ref{def Pinfty}) that, for $\lambda\in\mathbb
C\setminus \overline{\mathcal U}$,
\begin{align}
& R^{(1)}(\lambda)=-Q f'(u^c)^{-1/2}\frac{1}{\lambda -
u^c}\begin{pmatrix}0 & 1\\0&0
\end{pmatrix},\label{eqR1}\\
& R^{(2)}(\lambda)=\dfrac{1}{2f'(u^c)(\lambda-u^c)}
\begin{pmatrix}U+Q^2 & 0\\ 0 & -U-Q^2
\end{pmatrix},\label{eqR2}
\end{align}
 where
\begin{align*}&Q=Q(\epsilon^{-6/7}g_1(u^c;x,t_m),\epsilon^{-4/7}g_2(u^c;t_m)),\\
&
U=U(\epsilon^{-6/7}g_1(u^c;x,t_m),\epsilon^{-4/7}g_2(u^c;t_m)).\end{align*}
For the matrix $R^{(3)}$, using (\ref{RHPR3}) and
(\ref{R1})-(\ref{R2}), one obtains
\begin{multline*}
R^{(3)}(\lambda)=\frac{R^{(1)}(u^c)}{\lambda-u^c}
\Res(\Delta^{(2)}(\lambda);u^c)+\frac{R^{(2)}(u^c)}{\lambda-u^c}
\Res(\Delta^{(1)}(\lambda);u^c)+\frac{1}{\lambda-u^c}
\Res(\Delta^{(3)}(\lambda);u^c)\\
+\frac{1}{(\lambda-u^c)^2}
\Res((\lambda-u^c)\Delta^{(3)}(\lambda);u^c),\qquad\mbox{for
$\lambda\in\mathbb C\setminus \overline{\mathcal U}$.}
\end{multline*}
We are interested only in the $12$-entry of $R_3$, so it is
sufficient to compute the entries in the first line of
$R^{(1)}(u^c)$ and $R^{(2)}(u^c)$:
\begin{align*}
&R_{11}^{(1)}(u^c)=0,\\&
R_{12}^{(1)}(u^c)=\left(\dfrac{\epsilon^{-6/7}g_1'(u^c) Q_X+\epsilon^{-4/7}g_2'(u^c) Q_T}{\sqrt{f'(u^c)}}-\dfrac{Qf''(u^c)}{4f'(u^c)^{\frac{3}{2}}}\right),\\
&R_{11}^{(2)}(u^c)=\dfrac{f''(u^c)}{4f'(u^c)^2}U-\dfrac{\epsilon^{-6/7}
g_1'(u^c)}{2f'(u^c)}U_X
-\dfrac{\epsilon^{-4/7} g_2'(u^c)}{2f'(u^c)}U_T,   \\
&R_{12}^{(2)}(u^c)=0.
\end{align*}
Consequently,
\begin{equation}
\label{R3} R_{12}^{(3)}(\lambda)=\frac{1}{\lambda-u^c}Z
+\dfrac{1}{(\lambda-u^c)^2}\dfrac{V-W}{2f'(u^c)^{\frac{3}{2}}},\qquad\mbox{
for $\lambda\in\mathbb C\setminus \overline{\mathcal U}$,}
\end{equation}
where
\begin{multline}
\label{W} Z=\dfrac{\epsilon^{-6/7}
g_1'(u^c)}{2f'(u^c)^{\frac{3}{2}}}\left(Q^2Q_X+U_XQ-UQ_X+(V-W)_X\right)\\
+\dfrac{\epsilon^{-4/7}g_2'(u^c)}{2f'(u^c)^{\frac{3}{2}}}\left(Q^2Q_T+U_TQ-UQ_T+(V-W)_T\right)\\
-\dfrac{f''(u^c)}{8f'(u^c)^{5/2}}(Q^3+UQ+3(V-W)),
\end{multline}
and
\begin{align*}&V=V(\epsilon^{-6/7}g_1(u^c;x,t_m),\epsilon^{-4/7}g_2(u^c;t_m)),\\
&
W=W(\epsilon^{-6/7}g_1(u^c;x,t_m),\epsilon^{-4/7}g_2(u^c;t_m)).\end{align*}

\medskip

Compatibility of the small $\epsilon$-expansion
(\ref{Rexpansionepsilon}) with the large $\lambda$-expansion
 (\ref{asRlambda})
learns us that
\begin{equation}\label{R1expansion}
R_{1,12}(x,t,\epsilon)=-\epsilon^{1/7}Q\,
f'(u^c)^{-1/2}+\epsilon^{3/7}Z+\bigO(\epsilon^{4/7}),\end{equation}
in the double scaling limit. Taking $x$-derivatives in
(\ref{formulaR1}), one justifies that we can formally differentiate
(\ref{R1expansion}) to obtain
\begin{equation}\label{R1expansion2}
\begin{split}
-2\e\partial_xR_{1,12}(x,t,\epsilon)&=2\epsilon^{2/7}\dfrac{ \partial_xg_1(u^c)}{
f'(u^c)^{1/2}}U-2\epsilon^{4/7}\partial_xg_1(u^c)Z_X\\
&-2\epsilon^{4/7}\dfrac{
\partial_xg_1'(u^c)}{2f'(u^c)^{\frac{3}{2}}}\left(Q^2Q_X+U_XQ-UQ_X+V_X-W_X\right)+\bigO(\epsilon^{5/7}),
\end{split}
\end{equation}
since $Q_X=U$. Now we obtain by (\ref{f1})-(\ref{g22}), the first
equality in (\ref{uRa}), and (\ref{W}) that
\begin{multline*}
u(x,t_m,\e) =u_c+\left(\dfrac{2\e^2}{k^2}\right)^{1/7}
U\,\\
\qquad -\epsilon^{4/7}\dfrac{f''(u^c)}{2(8k)^{6/7}}\left( QU_X+U_{XX}+4U^2-3U_T[\e^{-6/7}g_1(u^c)]\right)\\
+\dfrac{g_2(u^c)}{(8k)^{4/7}}\left(\dfrac{2(m-1)}{5u^c}-\dfrac{3}{4}\dfrac{f''}{f'}\right)(2U_XQ_T+4UU_T+\dfrac{1}{2}U_{XXT})+
\bigO(\epsilon^{5/7}).
\end{multline*}
In the derivation of this expansion we have used
(\ref{Q})-(\ref{V}). For the third term, we also used the identity
(\ref{UKdV}). The term in the third line is also of order
$\bigO(\e^{4/7})$ because $g_2(u^c)$ is of order $\bigO(\e^{4/7})$.
The expansion for $u(x,t_m,\e)$ can be written in terms of $U$
exclusively by substituting the formula (\ref{Q}) for $Q$ and
$Q_T=-U^2/2-U_{XX}/12$. Substituting the values for $f''(u^c)$,
$g_1(u^c)$, and $g_2(u^c)$ given in (\ref{f1}), (\ref{g11}), and
(\ref{g21}), we find (\ref{expansionu}), with the constants $c_1,
c_2, c_3$ given by (\ref{c1})-(\ref{c3}). This proves Theorem
\ref{theorem: 2}.

\section*{Acknowledgements}
This work has been supported by the project FroM-PDE
funded by the European Research Council through the Advanced Investigator Grant
Scheme. TC also acknowledges support by the Belgian Interuniversity Attraction Pole
P06/02.

\obeylines \texttt{Tom Claeys
    Universit\'e Catholique de Louvain
Chemin du cyclotron 2
1348 Louvain-La-Neuve, BELGIUM
E-mail: tom.claeys@uclouvain.be
\bigskip

Tamara Grava
    SISSA
    Via Bonomea 265
    34136 Trieste,    ITALY
    E-mail: grava@sissa.it }

\end{document}